\documentclass[sigconf]{acmart}
\AtBeginDocument{%
  }

\usepackage{graphicx} 
\usepackage{multirow}
\usepackage{caption,subcaption}
\usepackage{color}
\usepackage{rotating}
\usepackage{tabularray}

\copyrightyear{2025}
\acmYear{2025}
\setcopyright{cc}
\setcctype{by}
\acmConference[CHI '25]{CHI Conference on Human Factors in Computing
Systems}{April 26-May 1, 2025}{Yokohama, Japan}
\acmBooktitle{CHI Conference on Human Factors in Computing Systems (CHI
'25), April 26-May 1, 2025, Yokohama,
Japan}\acmDOI{10.1145/3706598.3713382}
\acmISBN{979-8-4007-1394-1/25/04}

\begin{document}

\title{Text Entry for XR Trove (TEXT): Collecting and Analyzing Techniques for Text Input in XR}


\author{Arpit Bhatia}
\email{arbh@di.ku.dk}
\orcid{0000-0003-2356-9198}
\affiliation{%
  \institution{University of Copenhagen}
  \city{Copenhagen}
  \country{Denmark}
}

\author{Moaaz Hudhud Mughrabi}
\email{moaaz@is.mpg.de}
\affiliation{%
  \institution{Arizona State University}
  \city{Tempe}
  \state{Arizona}
  \country{United States}
}
\affiliation{%
  \institution{MPI for Intelligent Systems}
  \city{Stuttgart}
  \country{Germany}
}

\author{Diar Abdlkarim}
\email{diarkarim@gmail.com}
\author{Massimiliano Di Luca}
\email{m.diluca@bham.ac.uk}
\affiliation{%
  \institution{University of Birmingham}
  \city{Birmingham}
  \country{United Kingdom}
}

\author{Mar Gonzalez-Franco}
\email{margon@google.com}
\affiliation{%
  \institution{Google}
  \country{United States}}

\author{Karan Ahuja}
\email{kahuja@northwestern.edu}
\affiliation{%
  \institution{Google}
  \country{United States}}
\affiliation{%
  \institution{Northwestern University}
   \city{Evanston}
  \state{Illinois}
  \country{United States}}

\author{Hasti Seifi}
\email{hasti.seifi@asu.edu}
\affiliation{%
  \institution{Arizona State University}
  \city{Tempe}
  \state{Arizona}
  \country{United States}
}

\renewcommand{\shortauthors}{Bhatia et al.}

\begin{abstract}
Text entry for extended reality (XR) is far from perfect, and a variety of text entry techniques (TETs) have been proposed to fit various contexts of use. However, comparing between TETs remains challenging due to the lack of a consolidated collection of techniques, and limited understanding of how interaction attributes of a technique (e.g., presence of visual feedback) impact user performance. To address these gaps, this paper examines the current landscape of XR TETs by creating a database of 176 different techniques. We analyze this database to highlight trends in the design of these techniques, the metrics used to evaluate them, and how various interaction attributes impact these metrics. We discuss implications for future techniques and present TEXT: Text Entry for XR Trove, an interactive online tool to navigate our database.

\end{abstract}


\begin{teaserfigure}
  \includegraphics[width=\textwidth]{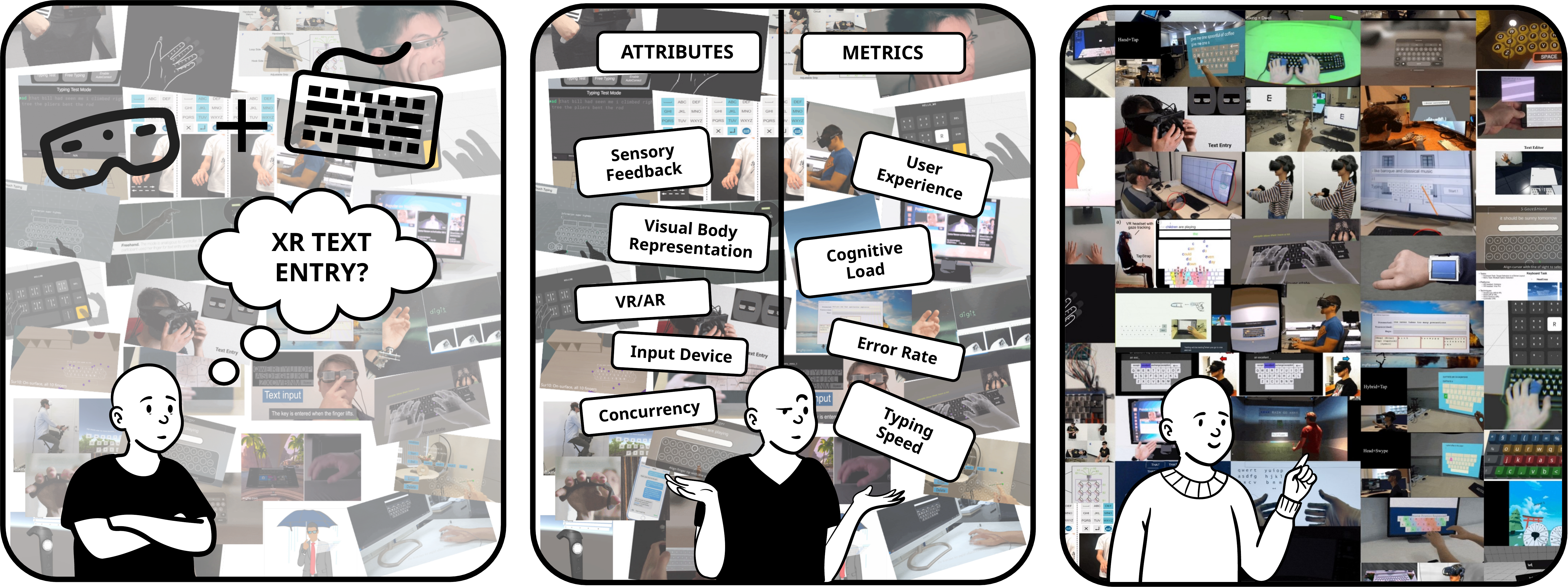}
  \caption{The space of text entry techniques for XR is large and fragmented (left) and XR designers are faced with many interaction attributes and little guidelines about how they impact user performance and experience (middle). We collect and analyze 176 techniques proposed in the last decade to support the selection and design of text entry techniques for XR applications (right).}
  \label{fig:teaser}
  \Description{Figure 1 shows a scattered collage of all the text entry techniques reviewed in this paper. Which when thought of in terms of the different attributes and metrics can become a useful tool for understanding text entry.}
\end{teaserfigure}

\received{20 February 2007}
\received[revised]{12 March 2009}
\received[accepted]{5 June 2009}

\maketitle

\section{Introduction}
Virtual, Mixed, and Augmented Reality (XR) technologies have the potential to integrate into our everyday lives and transform how we perform tasks, similar to how smartphones and personal computers did in the past. Unfortunately, text entry is one area where XR environments lag behind personal computers and mobile devices. Traditional text input methods suffer from inferior performance in XR due to issues such as tracking accuracy, stereo deficiencies, display resolution, and spatial awareness. Without proper text input methods in XR, the development of productivity tools, immersive \textit{metaverse} experiences, and potential \textit{killer} apps for extended productivity remains hindered \cite{gonzalez2024guidelines}. 

While a large number of text entry techniques (TETs) have been proposed for XR, there is no silver bullet solution. 
The unique challenges of XR environments necessitate tailored approaches. Yet, finding an appropriate technique from the ever-increasing list of techniques is not an easy task for several reasons. 
First, the descriptions of the different TETs are fragmented across academic research, applications from industry, or prototypes from hobbyists. 
Second, there is no repository of standardized performance metrics that can be achieved with each TET. 
Third, the TETs are seldom compared in terms of features, shortcomings, and advantages. 
Several categorizations of TETs exist \cite{dube_text_2019, 10.1145/3173574.3174221}, but a single level of categorization is unable to adequately capture the variety of techniques and the diversity of designers' needs for their applications. 
No comprehensive collection of TETs exists that provides a ``big picture'' of TETs to allow researchers and industry practitioners to navigate the solution space of text entry in XR.

As a result, selecting and designing TETs is often left to the intuition and comparison attempts of XR developers and researchers for each XR application. Only a limited number of works have provided guidelines to select a technique \cite{gonzalez_evaluation_2009, 10.1145/3173574.3174221}, but the choices are only based on a small set of related techniques. For other technologies, interactive collections have helped capture the evolution of fields and offered flexibility in browsing and comparing techniques~\cite{10.1145/3411764.3445319,seifi2019haptipedia,aigner2007visualizing,tominski2017timeviz,7156366,10.1145/3580592}. Yet, no such tool exists for XR TETs. 
Also, for those who wish to contribute by adding new techniques to the solution space, little is known about how different interaction attributes (e.g., presence of haptic or visual feedback) affect performance metrics. While individual studies have examined the impact of specific factors \cite{8446250, 8446059}, the links between attributes and metrics are still unclear across diverse techniques.

To address these gaps and by taking inspiration from previous interactive databases, 
this work collects and analyzes a set of 176 TETs for XR from various sources and presents an interactive visual database of the techniques with their attributes and metrics (Figure~\ref{fig:teaser}). We identify interaction attributes relevant to describing a technique and code each technique using them. Based on this data, we analyze trends in the interaction attributes and performance metrics used to evaluate these techniques. Finally, we analyze the impact of interaction attributes on the text entry performance. Based on our findings, we then provide a list of recommendations for future research on developing text entry techniques for XR.


The contributions of this work are the following:
\begin{itemize}
    \item A comprehensive database of 176 TETs for XR, each characterized according to 32 codes including 13 interaction attributes and its reported performance metrics.
    \item An analysis of trends in interaction attributes and their impact on the performance of TETs for XR. 
    \item TEXT: an online tool to navigate the solution space, consisting of a visualization of each TET.
    
\end{itemize}
To support XR research and design in this area, we make the database and the associated tool available on the Text Entry for XR Trove (TEXT) website:  \textit{\href{https://xrtexttrove.github.io/}{https://xrtexttrove.github.io/}}

\section{Related Work}
The trove is based on the analysis of previous scientific works to give a unified view of the landscape of TETs. In this section, we summarize how text entry in XR has been studied in the past, what metrics are used to evaluate text entry solutions and findings from previous works about how the design of a text entry technique in XR influences its performance.

\subsection{Analyzing Text Input in XR}
One of the earliest works analyzing multiple text entry techniques for XR was by  \citet{doi:10.1177/154193120204602611} who drew techniques from mobile and wearable computing and compared their performance in VR. 
At that time, the authors found that no technique was acceptable in terms of performance, usability and user satisfaction. 
In a similar study, \citet{gonzalez_evaluation_2009} compared six TETs in XR and found that a physical mobile phone keyboard performed best. 
Based on their findings, they also created a decision tree to choose an input technique based on the usage scenario. \citet{boletsis_controller-based_2019} investigated controller-based methods and compared four popular techniques against each other. 
They explored metrics beyond just accuracy and speed and advocated for more comparative studies using multiple user experience metrics. 
Our work also compares TETs in XR but instead of conducting a comparative study of a small set of techniques, we analyze the trends in reported data across a large set of them.

Researchers have also surveyed existing techniques to better understand the design space of text input in XR.
\citet{dube_text_2019} reviewed 32 techniques and classified them based on input mechanisms into 11 categories, highlighting the advantages and disadvantages of each category of techniques. 
Our work, instead of only having only one level of categorization, describes techniques using multiple attributes and highlights the most important ones. 
\citet{lewis2023virtual} reviewed typing techniques in XR to come up with a more modern technique and proposed to use speech for text entry in XR by comparing the performance of pure speech against speech plus a drum keyboard. 
\citet{10.1145/3173574.3174221} analyzed techniques that work by selecting characters on a keyboard. They proposed a design space for selection-based text entry in VR and also conducted a study comparing six such methods, providing a set of design guidelines for text entry in VR.

Our work expands on the idea of a design space for text entry in XR by considering the widest range of text entry techniques compared to prior work. 
For this, we not only review scientific works, but we expand our search to include unpublished TETs and non-academic sources (for details see Method section).


\subsection{Text Input Evaluation}
Prior work has established a set of versatile and repeatable evaluation methods that include performance metrics and user experience evaluations for comparing and optimizing text entry techniques. To test a technique, designers ask a diverse group of users to enter a specific set of phrases while measuring the text entry performance (e.g. speed of the entry, error rate) and then enquire about the user experience depending on the application.

\subsubsection{Phrase Sets}
When evaluating a technique, designers should make sure that the set of text phrases creates minimal bias on the text input. \citet{MacKenzieSoukoreffPhraseSet2003} argued for standardized phrase sets to enable repeatability and comparison in TET evaluations and introduced a 500-phrase set with strong internal and external validity \textit{(Mackenzie phrase set)}. 
\citet{PaekHsu2011Ngram} expanded on the notion of standardization by introducing a method to sample a phrase set from any corpus based on the notion of representativeness from information theory. 
This approach 
enables the creation of large representative phrase sets, thereby improving external validity and facilitating longitudinal studies. \citet{EnronPhraseSet2011} present the \textit{(Enron phrase set)} from mobile emails by Enron employees to ensure the phrases held semantic meaning and improve memorability because the phrases are real sentences written by everyday mobile device users. \citet{10.1145/3290605.3300821} created a challenging phrase set based on Twitter messages containing out-of-vocabulary words that are hard for a decoder to infer and can lead to auto-correction errors. In a review, \citet{KristenssonVertanen2012Comparison} compared different phrase sets and 
recommended using \textit{Mackenzie} or \textit{Enron} depending on whether the entry is on mobile devices. Given their importance, we extract which phrase set(s) designers used to test their techniques.

\subsubsection{Performance Metrics}
Typing speed is an important metric for evaluating TETs that the designers aim to maximize. The most common measure of text entry is \textit{words per minute} (WPM) \cite{Yamada1981Keyboards, ArifStuerzlinger2009}. $ WPM = \frac{|T|-1}{S} \times 60 \times \frac{1}{5}$, where $T$ is the typed text and $S$ is the total entry time, including backspaces, edits, etc. Other metrics can be used to characterize the speed of entry, such as adjusted word per minute, keystrokes per second, gestures per second \cite{ArifStuerzlinger2009}, or character per second \cite{Giovannelli2022intermittentTyping}. 
In this work, we collected WPM for typing speed as the commonly reported speed metric in the literature.

Quantifying errors in text entry allows one to capture the trade-off between speed and accuracy. \citet{SoukoreffMackenzie2003ErrorMetrics} define several issues with previous metrics for error rate and present \textit{total, uncorrected, and corrected error rates} as a holistic set of metrics for comparison across devices, keyboard layouts, and study designs. \textit{Total error rate} (TER) = \textit{corrected error rate} (CER) + \textit{uncorrected error rate} (UER). $CER = \frac{IF}{C + INF + IF} \times 100\%$ is the ratio of the fixed errors ($IF$) to the correct ($C$), incorrect fixed ($IF$), and incorrect not fixed ($INF$). Likewise, $UER = \frac{INF}{C + INF + IF} \times 100\%$ is the ratio of the not fixed errors. Others still use \textit{minimum string difference error rate} (MSD ER)--sometimes called \textit{character error rate} (CER)--quantifies the error in terms of the ``fixing'' operations needed for the input text to become the desired text normalized by the text length \cite{SoukoreffMackenzie2003MSDER}. Since MSD ER ignores the edit operations by the user, it cannot capture corrections in studies where the participants are asked to correct their errors \cite{SoukoreffMackenzie2003ErrorMetrics}. Character or word-level error rates can make more sense depending on the technique's input level. Swipe-like or prediction-supported TET can also use the IF, INF, and C words to calculate the word-level TER or use \textit{minimum word distance} (MWD) that counts the number of word ``fixing '' operations. Less frequent metrics are found in the literature such as \textit{error rate} (ER) and number of backspace uses. In this review, we extract each technique's reported error metric to determine whether the calculation was done at the character or word level.

\subsubsection{User Experience}

Besides performance, human factors and user experience are also important when designing text entry techniques. Task workload, often measured using the \textit{NASA task load index} (NASA TLX) questionnaire \cite{HART1988139} 
includes six factors capturing the user perception of mental, physical, and temporal demand of interaction as well as their perceived performance, effort, and frustration with the interaction. \textit{System usability scale} (SUS) \cite{brooke1996SUS} can quantify the ease of use, efficiency, and effectiveness of new interactions or devices for text entry. Researchers have also used custom ratings (e.g., preference) and statements to capture user experience. We extract NASA TLX scores when available and report a list of other experience metrics reported in TET studies.

\subsection{Factors Affecting Text Input in XR}
To assess the importance of interaction attributes for text entry, researchers vary the attributes in an interaction technique and compare the performance of these variants through a study. Investigating the representation of the user's hands,  \citet{8446250} compared the performance of showing no hands, avatar hands, fingertips only, and video of the hands for typing in VR. They reported no statistically significant effects on speed but found that fingertips and video of the hands had lower error rates. In a similar study,  \citet{10.1145/3173574.3173919} investigated various hand representations, such as displaying the hand skeleton and various levels of transparency for the visualizations. They again found no effects on speed but found that inexperienced typists benefited from having hand visualizations to orient themselves. Realistic hands also led to the highest presence and lowest workloads. \citet{10.1145/2702123.2702382} investigated the view of the keyboard available to typists by comparing typing performance in reality, augmented reality, and virtual reality. They found visual feedback essential to preserve typing performance in XR and proposed the augmented reality condition as a viable solution for typing with a keyboard in XR.

The visualization of the keyboard has been found to be a major factor that contributes to typing performance. \citet{10.1145/3334480.3382882} found key shapes to impact entry speed and dimension impacting accuracy along with both these attributes impacting user experience. \citet{yildirim_text_2020} compared flat vs. curved keyboards revealing that 2D keyboards led to higher speed and fewer corrections. \citet{8446059} compared the effect of repositioning the keyboard and hands in front of the user instead of at hand level and found no effect on typing speed, error rate, and NASA TLX scores when changing the position for using a physical keyboard. However, a drop in typing speed was observed when using a touchscreen keyboard. Touchscreen keyboards also led to significantly slower typing speeds than physical keyboards in XR. 

Feedback on typing is another influencing factor for XR text entry. \citet{10.1145/3025453.3025783} found the presence of visual feedback on key presses to improve error rates when typing on a visually occluded physical keyboard. \citet{gupta2020investigating} explored haptic feedback for mid-air text entry at various locations on the hand. They found that the presence of vibrotactile feedback did not significantly impact speed or accuracy but led to lower mental demand and effort along with higher user preference. For delivering the feedback, the finger base was found to be the most ideal position. They also discovered that feedback on hovering, not just key activation was useful for reducing errors in the case of both haptic and audio-visual feedback. Further demonstrating the importance of hovering feedback, \citet{yildirim2023point} showed that haptic feedback on hovering led to higher text entry speed, and any form of hovering feedback leads to higher accuracy compared to no feedback. In a second study, the effect of feedback on key activation was studied which found that audio or haptic key activation feedback led to increased entry speed and visual key activation feedback led to higher accuracy.

While these studies demonstrate what factors can impact text entry in XR, the relative importance of these factors is unknown at a larger scale. Our work analyzes trends across a diverse set of techniques to identify the most important of these factors.

\section{Method}

\begin{figure}[t]
    \centering
    \includegraphics[width=0.5\textwidth]{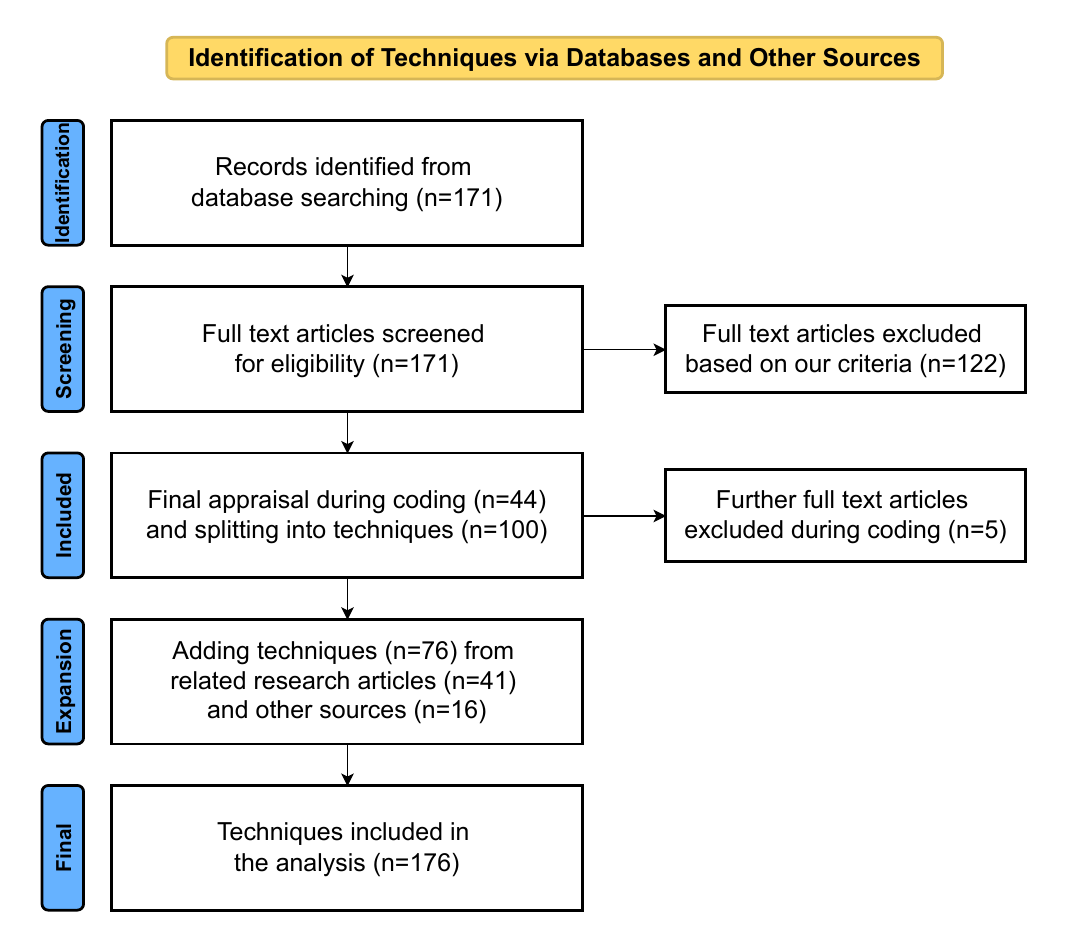}
    \caption{The five stages we used to identify relevant TETs for our review. These stages are inspired by the PRISMA Flow Diagram~\cite{PRISMA} and include the number of papers involved in each stage.}
    \label{fig:lit_rev}
    \Description{Figure 2 shows a 5-stage Identification of Techniques via Databases and Other Sources. In stage 1 171 Records are identified from database searching. In Stage 2 the records are screened for eligibility and results with 122 being excluded based on our criteria and 44 going to Stage 3. In Stage 3 44 of the included records were finally appraised and split into 100 techniques and 5 records were excluded during coding. In stage 4, 76 new techniques are added from 41 research articles and 16 other sources to expand the pool. The final stage included 176 techniques in the analysis.}
\end{figure}

In this section, we describe the process for identifying, screening, and analyzing relevant techniques for our database.
We first created an initial database of papers based on a systematic search of academic research articles on typing in XR based on the PRISMA Flow Diagram~\cite{PRISMA} for systematic reviews. 
We then expanded this database by considering techniques from non-academic sources such as commercial applications, social media, and blog posts. Additionally, we added more academic research articles that may not have been captured by our systematic search, for example, articles not published in XR-specific venues, or general TETs that are not developed with XR in focus but have been applied to XR (see Figure~\ref{fig:lit_rev} for complete process).

\subsection{Identification}

To begin our search for papers related to text input for XR, we searched through seven leading venues on XR and HCI: the ACM Conference on Human Factors in Computing Systems (CHI), ACM; the ACM Symposium on User Interface Software and Technology (UIST), ACM; the ACM Symposium on Virtual Reality Software and Technology (VRST), ACM; the IEEE Conference on Virtual Reality and 3D User Interfaces (IEEE VR), IEEE; IEEE Transactions on Visualization and Computer Graphics (TVCG), IEEE; the IEEE International Symposium on Mixed and Augmented Reality (ISMAR) , IEEE; and Virtual Reality journal (VR), Springer. We decided to focus our search on papers published after 2012 as it was the year the first modern VR headset was introduced by Oculus. We used the following query to search in the title and abstract of every publication in the selected venues: \\

\noindent \texttt{(text OR typing OR keyboard) AND (virtual reality OR VR OR augmented reality OR AR OR mixed reality OR MR OR extended reality OR XR OR HMD)} \newline

With our focus on capturing papers focused on typing in XR, the first portion of the query contains keywords related to text input and the second portion contains keywords related to XR. 
For the ACM and IEEE venues, we directly used the search options provided by the ACM Digital Library and IEEE Xplore respectively. 
For the VR journal, Springer only supports searching on the full text of the paper. 
Therefore, we used a custom script to further filter the results from the Springer website, ensuring that the search terms appeared in the title or abstract. 
We ran our query on January 2024 and thus the results include full papers published until 2023 covering the past 11 years (2013-2023, both included). 
This query resulted in 171 papers, 25 from CHI, 3 from UIST, 19 from VRST, 37 from IEEE VR, 36 from TVCG, 36 from ISMAR and 15 from VR. 

\subsection{Screening}
The results from the search query were then screened to only extract papers that either proposed a new TET or compared existing ones. Papers on reading text on XR headsets~\cite{10.1145/3411764.3445606} or editing techniques~\cite{9974482} were excluded based on this criteria. This was done as the goal of such interactions and the interaction techniques used were different from text entry.

Each paper was marked for inclusion (using 0 for excluded and 1 for included) by two authors who skimmed through the full-text article. The overall agreement percentage was 98.24\%, and Cohen's Kappa was 0.96. Each discrepancy in rating was discussed among all authors. In this phase, 122 papers were excluded resulting in a sample of 49 papers. During coding, 5 more papers were removed due to not meeting our inclusion criteria leading to a final set of 44 papers. 
 
\subsection{Expanding}
Having created an initial database of text input papers in XR from academic sources, we then proceeded to expand our database by collecting techniques from non-academic sources as well as more academic articles. For non-academic sources, the authors added techniques to the database through snowball search by adding techniques they were already aware of and then searching for more techniques using the keywords mentioned in the websites or posts they sourced the techniques from (e.g. Reddit, Dribbble, Medium). To find more relevant academic papers, we used the tool ResearchRabbit\footnote{https://www.researchrabbit.ai/} to look through papers and extended abstracts that are cited by the papers included in our systematic search. This resulted in an addition of 62 techniques from non-academic sources as well as 14 additional techniques from academia. 

\subsection{Splitting Papers Into Techniques}

After having finished adding new papers to our database, we split the papers into multiple techniques if they included more than one technique. The method for deciding how to split the papers was the following:

\begin{itemize}
    \item If a paper contained variations of a new TET (e.g., changing visual representation of the hands) and the authors compared them to propose a final novel technique (e.g., \cite{9874256, 8642443}), only the final version of the technique was included. i.e., we did not split the paper into multiple techniques.
    \item If a paper contained multiple TETs that the authors were comparing, the paper was split and all techniques were individually coded. For example, \cite{8943748, 9089533} were divided into 2 and 8 techniques, respectively.
\end{itemize}

\subsection{Coming Up With Attributes and Coding}

To analyze the techniques, we created a coding scheme that could be used to describe a TET. We looked at previous categorizations \cite{dube_text_2019, 10.1145/3173574.3174221}, skimmed through various research articles describing techniques (during the screening phase), watched videos of some of the non-academic techniques together as a group, and used our own prior experience in XR and text entry to come up with a list of 12 initial codes. These codes included interaction attributes (e.g., input device) and free text descriptions of the outcome measures (e.g., speed). We did not code for the technology being used itself but the interaction attributes they enable as technologies evolve rapidly over time and any technology-specific analysis would be outdated very quickly.

Based on these codes, the first round of coding took place where five of the authors coded 10 techniques and made notes for possible changes to the set of codes. In a joint meeting, the authors then discussed their notes to create an updated set of 32 codes which came out of trying to capture as much variation in the interaction attributes (13 codes) and splitting the outcome measures descriptions into separate codes (14 codes). We also included the metadata such as the source for each technique (5 codes).  We then created new descriptions of the codes (i.e., a codebook) and re-coded the 10 techniques based on the revised codebook. After this phase, the inter-rater reliability was at least 75\% for each code. Any differences were discussed in a meeting to agree on the final codes for each technique. 

The remaining TETs were divided so that each of them was coded by one of six authors. A seventh author who joined later first coded the same 10 techniques from the previous round and then compared their codes with the final codes to check for errors and only then moved to code new techniques. 

\subsection{Attributes and Metrics}
Each technique in our database is described using 32 codes. Among these, 13 codes correspond to the interaction attributes (Table \ref{tab:attributes}), and 14 correspond to performance metrics that are frequently reported by studies (see Table \ref{tab:metrics} for the top six reported metrics). The other five codes provide additional information on the source of the technique (academia/industry/hobbyists), date of release, presence of study, phrase set used, and other metrics reported beyond those captured by our metrics codes. We provide the full codebook with the description of attributes in the supplementary materials. 

\definecolor{AthensGray}{rgb}{0.921,0.921,0.925}
\definecolor{Shark}{rgb}{0.098,0.098,0.101}
\begin{table*}
\centering
\caption{Attributes and their corresponding values used to describe a TET (the percentages may not sum to 100\% as one attribute can have multiple values).}
\label{tab:attributes}
\resizebox{\linewidth}{!}{%
\begin{tabular}{l|lllllll}
\textbf{\LARGE Attributes} & \textbf{\LARGE Values and Examples} &  &  &  &  &  & \\
\hline
\textbf{\large Input Device} & \textbf{None} \cite{8943750} & \textbf{Custom} & \textbf{Physical~} & \textbf{Controller} \cite{10.1145/3385959.3421722}& \textbf{Touchscreen} & \textbf{Other} \cite{10.1145/3491102.3501838} & \\
\textit{(The hardware that~} &  & \textbf{Hardware} \cite{10.1145/3290605.3300244} & \textbf{Keyboard} \cite{8699306}&  & \cite{10.1145/3173574.3174102} &  & \\
\textit{is used to input the letters)} & 56 \textcolor[rgb]{0.6,0.6,0.6}{(31.82\%)} & 46 \textcolor[rgb]{0.6,0.6,0.6}{(26.14\%)} & 37 \textcolor[rgb]{0.6,0.6,0.6}{(21.02\%)} & 33 \textcolor[rgb]{0.6,0.6,0.6}{(18.75\%)} & 10 \textcolor[rgb]{0.6,0.6,0.6}{(5.68\%)} & 3 \textcolor[rgb]{0.6,0.6,0.6}{(1.7\%)} & \\
\hline
\textbf{\large Body Part (for input)} & \textbf{Finger(s) \cite{10.1145/3432204}} & \textbf{Hand(s)} \cite{technologies7020031}& \textbf{Head} \cite{10.1145/3380983}& \textbf{Gaze} \cite{Rajanna2018onthegostudy1}& \textbf{Voice} \cite{9382836}&  & \\
\textit{(The body part used for entering text)} &  &  &  &  &  &  & \\
 & 132 \textcolor[rgb]{0.6,0.6,0.6}{(75.0\%)} & 54 \textcolor[rgb]{0.6,0.6,0.6}{(30.68\%)} & 15 \textcolor[rgb]{0.6,0.6,0.6}{(8.52\%)} & 14 \textcolor[rgb]{0.6,0.6,0.6}{(7.95\%)} & 1 \textcolor[rgb]{0.6,0.6,0.6}{(0.57\%)} &  & \\
\hline
 \textbf{\large Concurrency} & \textbf{One} \cite{10.1145/3332165.3347865}& \textbf{Multiple} \cite{8797740}& \textbf{Two} \cite{10.1145/3388770.3407399}&  &  &  & \\
\textit{(How many pointers does the user~have~} &  &  &  &  &  &  & \\
\textit{in the virtual environment for text entry.)} & 83 \textcolor[rgb]{0.6,0.6,0.6}{(47.16\%)} & 61 \textcolor[rgb]{0.6,0.6,0.6}{(34.66\%)} & 32 \textcolor[rgb]{0.6,0.6,0.6}{(18.18\%)} &  &  &  & \\
\hline
\textbf{\large Haptic Feedback Modality} & \textbf{Button~} & \textbf{None} \cite{10269033}& \textbf{On-body} \cite{10.1145/2785830.2785886}& \textbf{External~} & \textbf{Vibrotactile} \cite{gupta2020investigating}& \textbf{Force~} & \textbf{Other} \cite{10.1145/3173574.3174221}\\
(\textit{Type of haptic feedback}) & \textbf{Press} \cite{Giovannelli2022intermittentTyping} &  &  & \textbf{Surface} \cite{8943750}&  & \textbf{Feedback} \cite{8456570}& \\
 & 59 \textcolor[rgb]{0.6,0.6,0.6}{(33.52\%)} & 54 \textcolor[rgb]{0.6,0.6,0.6}{(30.68\%)} & 29 \textcolor[rgb]{0.6,0.6,0.6}{(16.48\%)} & 24 \textcolor[rgb]{0.6,0.6,0.6}{(13.64\%)} & 10 \textcolor[rgb]{0.6,0.6,0.6}{(5.68\%)} & 4 \textcolor[rgb]{0.6,0.6,0.6}{(2.27\%)} & 2 \textcolor[rgb]{0.6,0.6,0.6}{(1.14\%)}\\
\hline
\textbf{\large Haptic Feedback} & \textbf{Key} & \textbf{None} \cite{10.1145/2807442.2807504}& \textbf{Hovering} \cite{10.1145/2984511.2984576}& \textbf{Other} \cite{10.1145/3491101.3519679}&  &  & \\
(\textit{For what events or user actions~} & \textbf{Activation} \cite{10.1145/3334480.3382888}&  &  &  &  &  & \\
\textit{haptic feedback is provided to users}) & 101 \textcolor[rgb]{0.6,0.6,0.6}{(57.39\%)} & 57 \textcolor[rgb]{0.6,0.6,0.6}{(32.39\%)} & 56 \textcolor[rgb]{0.6,0.6,0.6}{(31.82\%)} & 11 \textcolor[rgb]{0.6,0.6,0.6}{(6.25\%)} &  &  & \\
\hline
\textbf{\large Audio Feedback} & \textbf{Key~} & \textbf{None} \cite{10.1145/3569463}& \textbf{Unknown} \cite{perfectvrkey} & \textbf{Hovering} \cite{pxrkeye}& \textbf{\textbf{Other}} \cite{10.1145/3025453.3025964}&  & \\
\textit{(For what events or user actions} & \textbf{Activation} \cite{10.1145/3385956.3422114}&  &  &  &  &  & \\
\textit{auditory~feedback is provided to users)} & 96~\textcolor[rgb]{0.6,0.6,0.6}{(54.54\%)} & 70 \textcolor[rgb]{0.6,0.6,0.6}{(39.77\%)} & 8 \textcolor[rgb]{0.6,0.6,0.6}{(4.55\%)} & 2 \textcolor[rgb]{0.6,0.6,0.6}{(1.14\%)} & 1~\textcolor[rgb]{0.6,0.6,0.6}{(1.14\%)} &  & \\
\hline
\textbf{\large Visual Feedback} & \textbf{Key~} & \textbf{Hovering} \cite{10.1145/3173574.3174102}& \textbf{None} \cite{10.1145/3379337.3415816}& \textbf{Other} \cite{10.1145/3569461}&  &  & \\
(\textit{For what events or user actions} & \textbf{Activation} \cite{8446250}&  &  &  &  &  & \\
\textit{visual~feedback is provided to users)} & 83 \textcolor[rgb]{0.6,0.6,0.6}{(47.15\%)} & 74 \textcolor[rgb]{0.6,0.6,0.6}{(42.04\%)} & 45~\textcolor[rgb]{0.6,0.6,0.6}{(25.56\%)} & 7 \textcolor[rgb]{0.6,0.6,0.6}{(}\textcolor[rgb]{0.6,0.6,0.6}{3.98\%}\textcolor[rgb]{0.6,0.6,0.6}{)} &  &  & \\
\hline
\textbf{\large Visual Body Representation} & \textbf{Full Hands} \cite{9417793} & \textbf{Cursor/} & \textbf{Invisible} \cite{10.1145/3385959.3418454}& \textbf{Fingertips} \cite{8446059}& \textbf{Controllers} \cite{9089533}& \textbf{Bones} \cite{8794572}& \textbf{Rays/}\\
(\textit{How is the user presented visually}) &  & \textbf{Pointer} \cite{10.1145/3379503.3403549}&  &  &  &  & \textbf{Drums} \cite{10.1145/3388770.3407399}\\
 & 59 \textcolor[rgb]{0.6,0.6,0.6}{(33.52\%)} & 43 \textcolor[rgb]{0.6,0.6,0.6}{(24.43\%)} & 38 \textcolor[rgb]{0.6,0.6,0.6}{(21.59\%)} & 16 \textcolor[rgb]{0.6,0.6,0.6}{(9.09\%)} & 11 \textcolor[rgb]{0.6,0.6,0.6}{(6.25\%)} & 11 \textcolor[rgb]{0.6,0.6,0.6}{(6.25\%)} & 8 \textcolor[rgb]{0.6,0.6,0.6}{(4.55\%)}\\
\hline
\textbf{\large Visual Keyboard Representation} & \textbf{Virtual~} & \textbf{Virtual~} & \textbf{None} \cite{10.1145/3472749.3474788}& \textbf{Passthrough} &  &  & \\
(\textit{How is the keyboard represented visually}) & \textbf{Keyboard} \cite{9284687}& \textbf{Overlay} \cite{8794572}&  & \cite{10233148} &  &  & \\
 & 112~\textcolor[rgb]{0.6,0.6,0.6}{(63.63\%)} & 32~\textcolor[rgb]{0.6,0.6,0.6}{(20.45\%)} & 22 \textcolor[rgb]{0.6,0.6,0.6}{(18.18\%)} & 10 \textcolor[rgb]{0.6,0.6,0.6}{(5.68\%)} &  &  & \\
\hline
\textbf{\large Keyboard Layout} & \textbf{QWERTY} \cite{10.1145/2702123.2702382}& \textbf{Other} \cite{montoya2023wordsphere} & \textbf{\textbf{Radial}} \cite{9874256}& \textbf{Adapted} & \textbf{Alphabetical} & \textbf{\textbf{QWERTZ}} \cite{10049665}& \textbf{None} \cite{10.1145/2785830.2785885}\\
\textit{(Order and layout of keys)} &  &  &  & \textbf{QWERTY} \cite{10.1145/3313831.3376306}&  \cite{Jiang17012024}&  & \\
 & 119 \textcolor[rgb]{0.6,0.6,0.6}{(67.61\%)} & 18 \textcolor[rgb]{0.6,0.6,0.6}{(10.23\%)} & 10~\textcolor[rgb]{0.6,0.6,0.6}{(}\textcolor[rgb]{0.6,0.6,0.6}{5.68\%}\textcolor[rgb]{0.6,0.6,0.6}{)} & 9 \textcolor[rgb]{0.6,0.6,0.6}{(5.11\%)} & 9 \textcolor[rgb]{0.6,0.6,0.6}{(5.11\%)} & 9~\textcolor[rgb]{0.6,0.6,0.6}{(5.11\%)} & 7 \textcolor[rgb]{0.6,0.6,0.6}{(3.98\%)}\\
\hline
\textbf{\large Keyboard Backend} & \textbf{None} \cite{8951915}& \textbf{Prediction} \cite{10.1145/3025453.3025964}& \textbf{Correction} \cite{10.1145/3025453.3025783}& \textbf{Personalization} \cite{9994904}&  &  & \\
\textit{(Algorithms that assist with text entry~by~} &  &  &  &  &  &  & \\
\textit{processing how typing is being performed)} & 98 \textcolor[rgb]{0.6,0.6,0.6}{(55.68\%)} & 67 \textcolor[rgb]{0.6,0.6,0.6}{(38.07\%)} & 23 \textcolor[rgb]{0.6,0.6,0.6}{(13.07\%)} & 4 \textcolor[rgb]{0.6,0.6,0.6}{(2.27\%)} &  &  & \\
\hline
\textbf{\large Can Be Mobile?} & \textbf{Stationary} \cite{10.1145/3173574.3174221}& \textbf{On-the-go} \cite{10.1145/3586183.3606803}&  &  &  &  & \\
(\textit{Can the user move around while typing}) &  &  &  &  &  &  & \\
 & 121~\textcolor[rgb]{0.6,0.6,0.6}{(68.75\%)} & 55~\textcolor[rgb]{0.6,0.6,0.6}{(31.25\%)} &  &  &  &  & \\
\hline
\textbf{\large VR/AR} & \textbf{VR only} \cite{8798238}& \textbf{AR only} \cite{10.1145/3611659.3615692}& \textbf{Both} \cite{10.1145/3491102.3501878}&  &  &  & \\
(\textit{For what XR technology~} &  &  &  &  &  &  & \\
\textit{was this technique~developed)} & 118 \textcolor[rgb]{0.6,0.6,0.6}{(67.05\%)} & 34 \textcolor[rgb]{0.6,0.6,0.6}{(19.32\%)} & 24 \textcolor[rgb]{0.6,0.6,0.6}{(13.64\%)} &  &  &  & \\
\bottomrule
\end{tabular}
}
\Description{Table 1 shows the attributes, their codes, and the number and percentage of techniques in each code.}
\end{table*}

For coding the metric values, we always use the average reported value for novice users. This is because the definition of experts varies across studies and very few techniques are evaluated using experts. For multi-session studies, we report values from the last study session to pass the initial learning curve and get the most accurate metric value. After populating all the metrics in our database, we further filled up missing values that we could by inferring values that are dependent on other reported metrics (e.g. TER based on CER and UER, Overall TLX based on the subscales). For analyzing the trends in metrics, we excluded during analysis the techniques that are only based on typing passwords \cite{9995649, 8794572} and those that only used expert participants with significant typing experience \cite{10.1145/3379337.3415816}.

\section{Text Entry for XR Trove (TEXT): An Online Tool}
Having coded all TETs, we created an online tool, \textbf{Text Entry for XR Trove (TEXT)} to help XR researchers and practitioners navigate our database. The tool is available at \textit{\href{https://xrtexttrove.github.io/}{https://xrtexttrove.github.io/}} and is based on the open source code from the \textit{Locomotion Vault} project \cite{10.1145/3411764.3445319}. The tool provides the following functionalities:
\begin{enumerate}
    \item A \textit{Gallery} of gifs or images of the techniques that allow users to see the techniques in action.
     \item A \textit{Detailed View} of each technique which displays all available information in our database about that particular technique. 
       \item A \textit{Filter panel}  to browse a subset of the techniques in our database based on specific attributes or performance metrics needed in an application. Our filter panel is based on OR logic and shows all techniques that satisfy a given filter. Thus, techniques with multiple values for a code (e.g., head and hand input modalities) appear when the filter for either category is selected. 
     \item A \textit{Suggestion Form} for adding new techniques and making modifications to the trove to keep it up to date.
\end{enumerate}

\begin{figure*}[!h]
    \includegraphics[width=\textwidth]{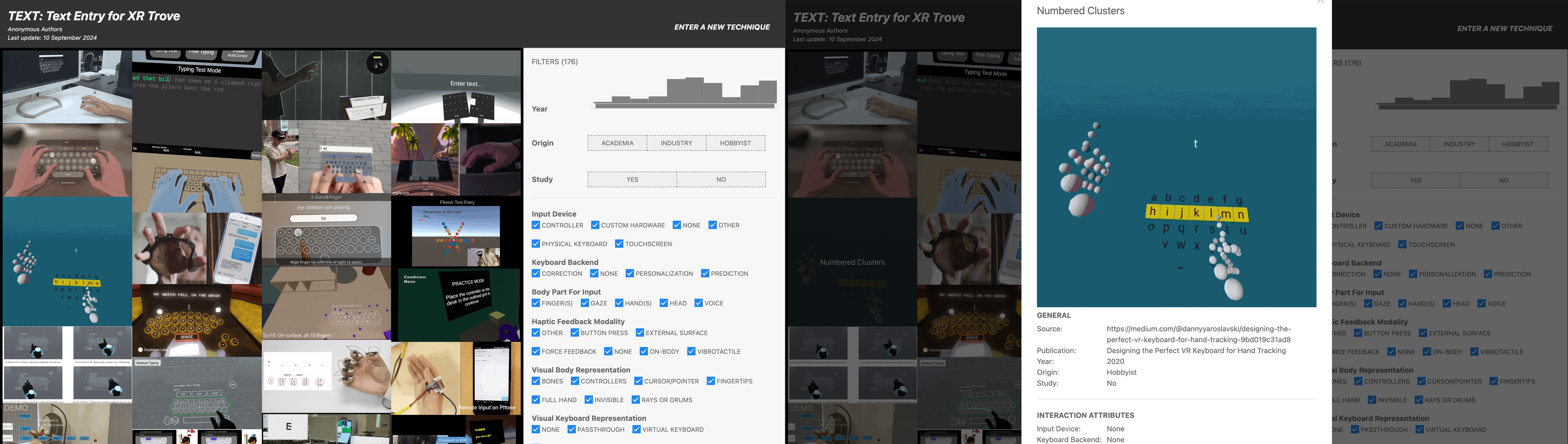}
    \caption{Screenshots of Text Entry for XR Trove (TEXT). On the left, the interactive \emph{gallery} of the TETs, a \emph{filter panel} for searching a subset of techniques, and a \emph{suggestion form} for reporting new techniques. On the right, a \emph{detailed view} of a technique with its metadata, interaction attributes, and performance metrics.}
    \label{fig:TEXT_Screenshots}
    \Description{Figure 3 Screenshots of Text Entry for XR Trove (TEXT). On the left, the interactive gallery of the TETs, a filter panel for searching a subset of techniques, and a suggestion form for reporting new techniques. On the right, a detailed view of a technique with its metadata, interaction attributes, and performance metrics}
\end{figure*}




\section{Results}
In this section, we describe what performance metrics are used for TETs, the relation between attributes and metrics, and trends in the design of TETs over time.

\subsection{Metrics for Text Entry in XR}

We found variations across TETs in reporting performance metrics. Techniques by hobbyists (N=17) never reported any performance or user experience metrics. Academia (N=151) and Industry (N=37) reported typing speed and error metrics for 90.7\% and 86.5\% of the techniques respectively. We present the reporting practices for each metric below.

\subsubsection{Typing Speed} Over 77.8\% (N=137) of the TETs reported WPM as the typing speed metric, making it the most common metric in the literature. Techniques without speed metric either lacked a study (N=32) or focused on factors such as learnability and key selection (N=2). 
Other metrics for speed were characters per minute (CPM) and ``perfect word'' typing speed \cite{10.1145/3334480.3382888} which excludes time spent on error correction.

\subsubsection{Accuracy} Unlike typing speed, accuracy metrics vary widely in the literature (Table \ref{tab:metrics}). TER was the most reported error metric in 58 TETs (33\%), followed by CER (31.8\%). However, they are usually reported together with UER (26.7\%), 8 of which are at the word level (4.5\%). MSD ER was reported as the error metric in 49 TETs (30.1\%), 5 of them were word-level errors (2.8\%). Other metrics like ER (4.5\%) were less common, while the rest of the techniques (32.9\%) either used unique error metrics or did not report any.

\subsubsection{NASA TLX Scores} Task workload was reported less frequently than speed and accuracy, with only 54 techniques reporting the overall workload (30.7\%). Of those, only 16 TETs reported all six subscales of the index. Some studies used five subscales, omitting \textit{Temporal Load} (N=2) or \textit{Performance} (N=9). Others only reported scales with significant differences, such as \textit{Mental-Physical} load (N=2) and \textit{Physical-Frustration} (N=6). 

\begin{table*}
\small
\begin{tabular}{lllllll}
\hline
\textbf{Metric} & \textbf{Words per minute} & \textbf{CER} & \textbf{UER} & \textbf{TER} & \textbf{MSD ER} & \textbf{Overall TLX} \\ \hline
\textbf{count} & 137 & 56 & 47 & 58 & 49 & 46 \\
\textbf{mean} & 20.36 & 5.99 & 2.13 & 7.96 & 3.77 & 0.48 \\
\textbf{min} & 2.8 & 0.5 & 0 & 0.5 & 0.5 & 0.24 \\
\textbf{max} & 55.6 & 30.26 & 15.1 & 30.87 & 23 & 0.8 \\ \hline
\end{tabular}
\caption{Summary Statistics for the Most Commonly Used Metrics}
\label{tab:metrics}
\Description{Table 2 shows that most techniques report words per minute as the speed metric, while error rate is split between Total error rate (or correct and uncorrected) and minimum string distance error rate. TLX scores are reported less in the reviewed techniques.}
\end{table*}

\begin{table*}
\small
\begin{tabular}{p{3.5cm}p{11cm}}
\hline
\textbf{Metric Category} & \textbf{Questionnaires and Frequency} \\ \hline
\textbf{General User Experience} & User Experience Questionnaire \cite{laugwitz2008construction} (N=26), System Usability Scale \cite{brooke2013sus} (N=11), Flow-Short-Scale \cite{rheinberg2003erfassung} (N=8), Borg CR10 scale \cite{borg1985increase} (N=5), Technology Acceptance Model 3 (N=3), Social Acceptability Questionnaire \cite{10.1145/2628363.2628381} (N=2), Game Experience Questionnaire \cite{ijsselsteijn2013game} (N=1), Custom (n=26).\\
\hline
\textbf{XR Experience} & Simulator Sickness Questionnaire \cite{kennedy1993simulator} (N=22), Motion Sickness Assessment Questionnaire \cite{gianaros2001questionnaire} (N=14), Slater-Usoh-Steed questionnaire (N=14), IPQ spatial presence questionnaire \cite{regenbrecht2002real} (N=10), presence questionnaire \cite{witmer1998measuring} (N=10).\\
\hline
\textbf{XR Text Entry and Motion} & Accuracy of first key press (N=8), Time to first key press \cite{10.1145/2702123.2702382} (N=22), Motion tracking measures (N=23), Statistics of non-typing actions (N=14).\\
\hline
\end{tabular}
\caption{Summary of Other Metrics Reported for TETs}
\label{tab:ux}
\Description{Table 3 divides the other metrics used in XR text entry studies into 3 categories (General user experience, XR experience, XR text entry and motion) and displays the metrics and the number of times they were reported.}
\end{table*}

\subsubsection{Other Metrics} Table~\ref{tab:ux} shows other user experience and movement metrics for TETs. Several TETs report general UX or XR-specific measures related to simulator sickness and sense of presence. 
Two metrics created for XR TETs are accuracy of and time to first key press (N=8 and N=22), which assess success in locating the input device \cite{10.1145/2702123.2702382}, though they are sometimes called homing time \cite{Giovannelli2022intermittentTyping} or time to first character \cite{10.1145/3359996.3364265}.
In some cases, motion tracking data (N=23) is also analyzed for typing behaviors, including press duration, finger travel, and number of finger-key collisions.
The influence of a technique on typing behavior is measured by statistics of non-typing actions (N=14) such as backspace usage, insertions, and auto-complete usage. Finally, for aspects of the techniques not covered by other metrics, interviews and open comments (N=30) are used.

\subsubsection{Relationship Between Metrics} We calculated the correlation between common metrics to assess their relationships. Pearson correlation was used for Overall TLX and TER, which met the bivariate normality assumption. For other metrics like WPM and TER, WPM and MSD ER, and TLX and MSD ER, which did not meet this assumption, we used Spearman rank correlation. Results showed a small positive correlation between WPM and TER ($\rho_s=0.32$, $p=0.015$) and a small negative correlation between WPM and MSD ER ($\rho_s=-0.29$, $p=0.045$). No other significant correlations were found ($p>0.05$), and plots did not reveal further interesting relationships (see Supplementary Materials).

\subsection{Identifying the Importance of Interaction Attributes for Typing Performance}

To identify and rank the importance of each attribute, we used Random Forest models to perform supervised feature selection. We chose Random Forests for their robustness to over-fitting and their ability to capture non-linear relationships in the data. This model also reduces the overlap of information between features, such as when the value of a feature inherently determines other features (e.g., gaze as the body part for input always having a concurrency of one), or how the absence of one feature (e.g., haptic feedback modality) dictates related features (e.g., haptic feedback). We also considered other feature selection techniques such as Decision Trees and Stepwise Regression but found similar results and less importance separation.

To assess feature importance, we use Gini importance \cite{menze2009comparison} as the split criterion in our Random Forest model. The Gini importance values range between 0 and 1 for each feature and they sum to 1 in a model. They quantify each feature's contribution to reduce data uncertainty, i.e. the feature's standalone information power \cite{saeys2008robust} in distinguishing between different TETs. In the results, we report attribute values with Gini importance over 5\% to focus on the most important features and also account for noise in statistical modeling (See Supplementary materials for all values). 

To create the models, we used 11 of our 13 interaction attributes. The two we excluded were `Can it be mobile?' and `VR/AR' because despite being applicable to the techniques, these attributes are not usually employed when evaluating them. 
Although many techniques were presented for on-the-go text entry, only 3 techniques \cite{Rajanna2018onthegostudy1, Lee2021onthegostudy2} conducted studies with the users actively moving while typing. 
For AR/VR, techniques applicable to both VR and AR are usually evaluated in only one of them and studies do not consider the impact of the environment on text entry. 
We also do not include `Phrase Set' in our analysis as it is a parameter for the user study rather than the technique itself. While we acknowledge that the phrase set can have an impact on typing performance, we focus on studying the impact of the interaction parameters. In our review, most studies (N=122 out of 140 TETs with studies) used one of the standardized corpora between which \citet{KristenssonVertanen2012Comparison} suggest that there is no statistically significant difference.

For the remaining attributes, we first converted our 11 categorical interaction attributes into binary representations using a one-hot-encoding scheme. This step created features for the model corresponding to each attribute value, such as Visual body representation: controllers, Haptic feedback: hovering, and so on. For each performance metric (e.g., words per minute, error rate, NASA TLX scores), we then built a Random Forests model based on these encoded attributes. This approach allows us to not only identify the most important attributes to focus on when designing new techniques but also to understand how different attributes contribute to specific performance metrics. For model training, we used a Random Forest regressor implemented through scikit-learn \cite{scikit-learn} in Python on Google Colab\footnote{https://colab.google/}. The models were trained using the default hyperparameters provided by scikit-learn. The training set for each model was the set of all techniques for which we had data for a given performance metric. Since we focus on identifying trends rather than using the model to predict a value, we do not split the data into training and test sets. 

\begin{figure*}[htb]
    \centering
    \begin{minipage}[t]{0.48\textwidth}
        \centering
        \begin{subfigure}[b]{\textwidth}
        \caption{}
            \includegraphics[width=\linewidth]{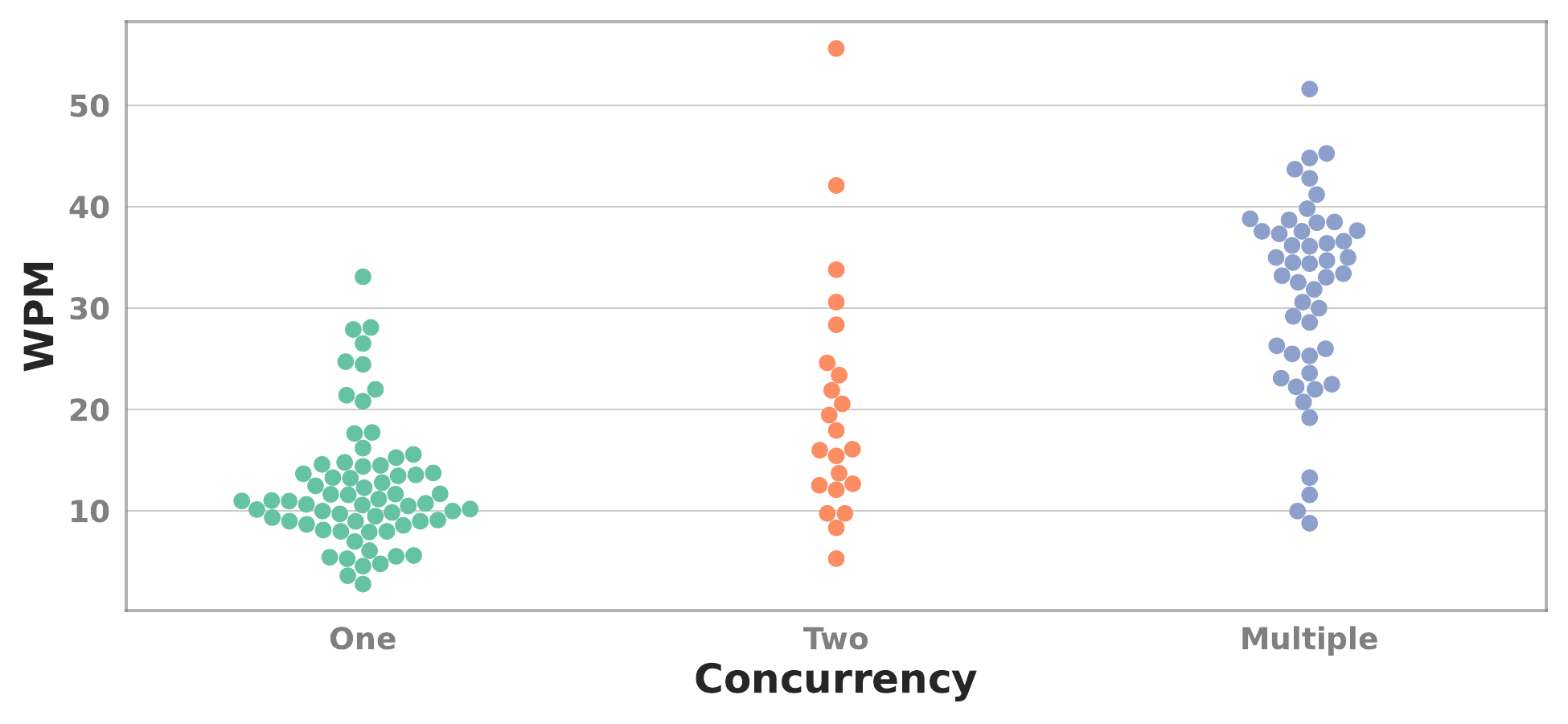}
            \label{fig:atribute_vs_metric:concurrency_WPM}
        \end{subfigure}
        \begin{subfigure}[b]{\textwidth}
        \caption{}
            \includegraphics[width=\linewidth]{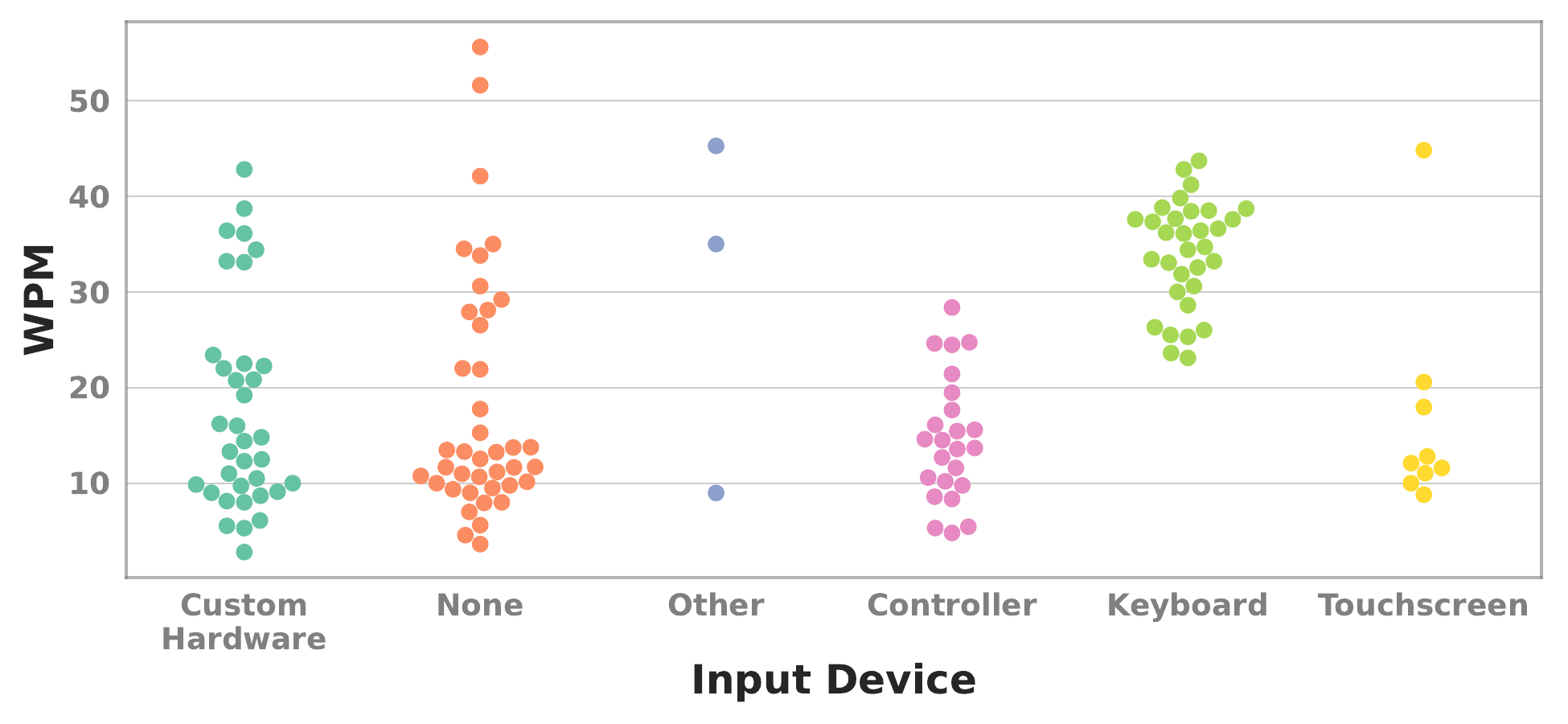}
            \label{fig:atribute_vs_metric:inputdevice_WPM}
        \end{subfigure}
        \begin{subfigure}[b]{\textwidth}
        \caption{}
            \includegraphics[width=\linewidth]{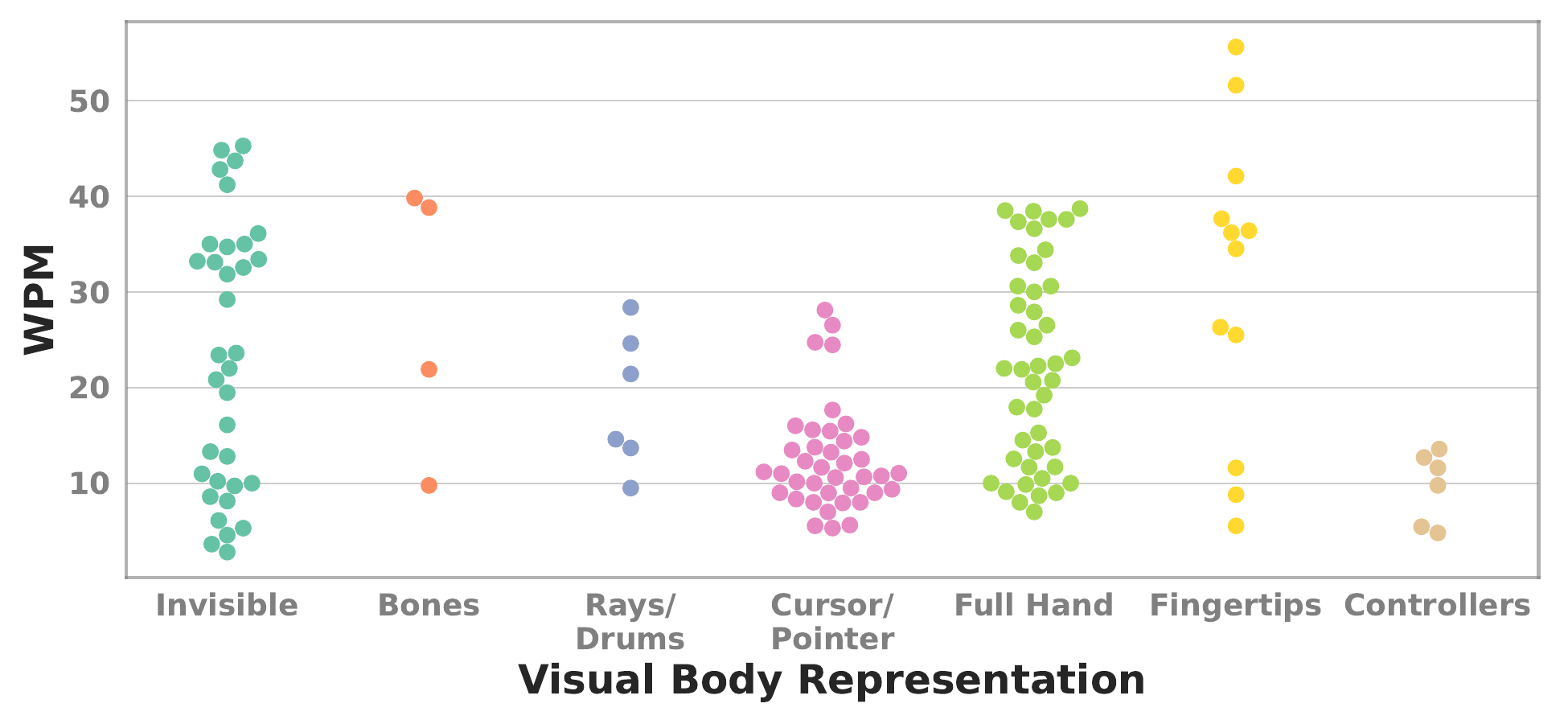}
            \label{fig:atribute_vs_metric:visbodyrep_WPM}
        \end{subfigure}
    \end{minipage}
    \hspace{2mm}
    \begin{minipage}[t]{0.48\textwidth} 
        \centering
        \begin{subfigure}[b]{\textwidth}
        \caption{}
            \includegraphics[width=\linewidth]{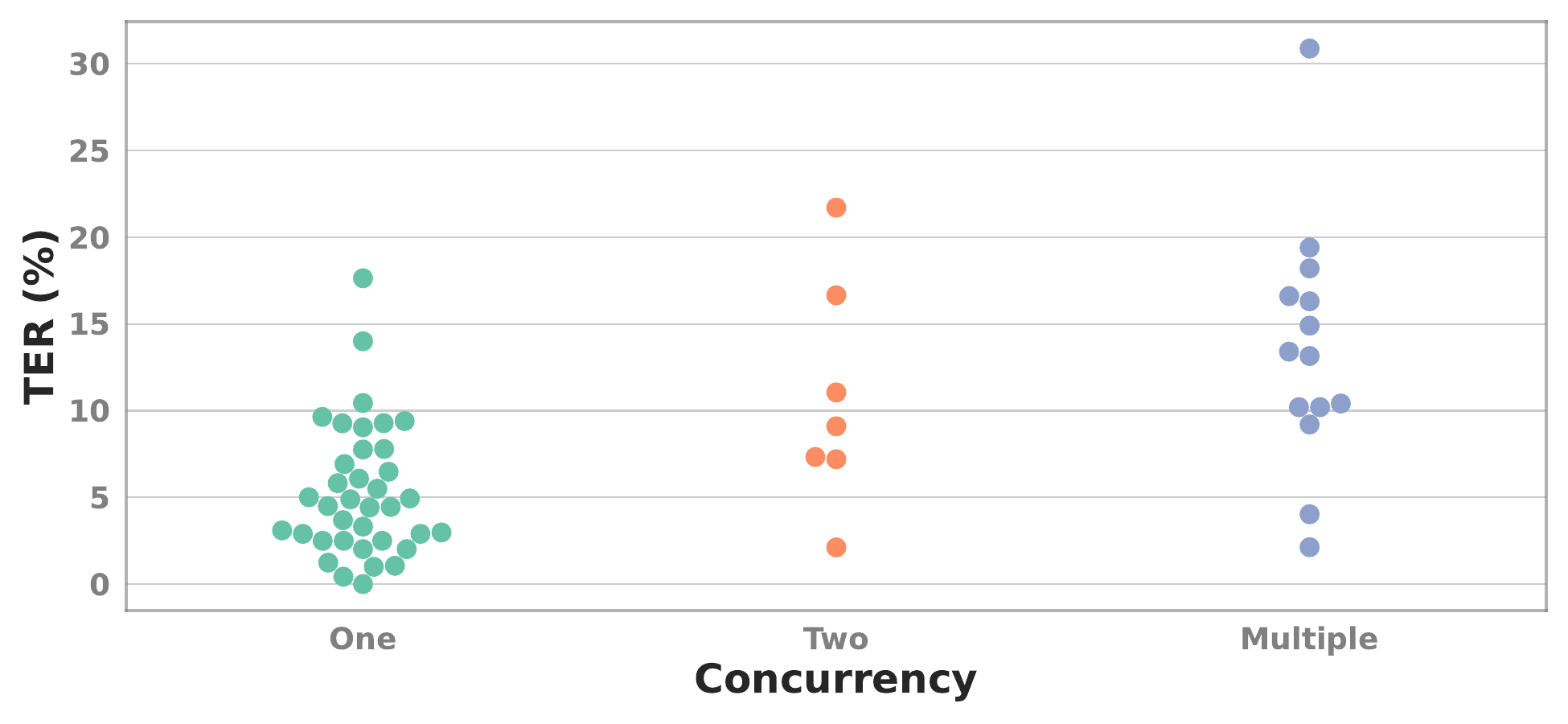}
            \label{fig:atribute_vs_metric:concurrency_TER}
        \end{subfigure}
        \begin{subfigure}[b]{\textwidth}
        \caption{}
            \includegraphics[width=\linewidth]{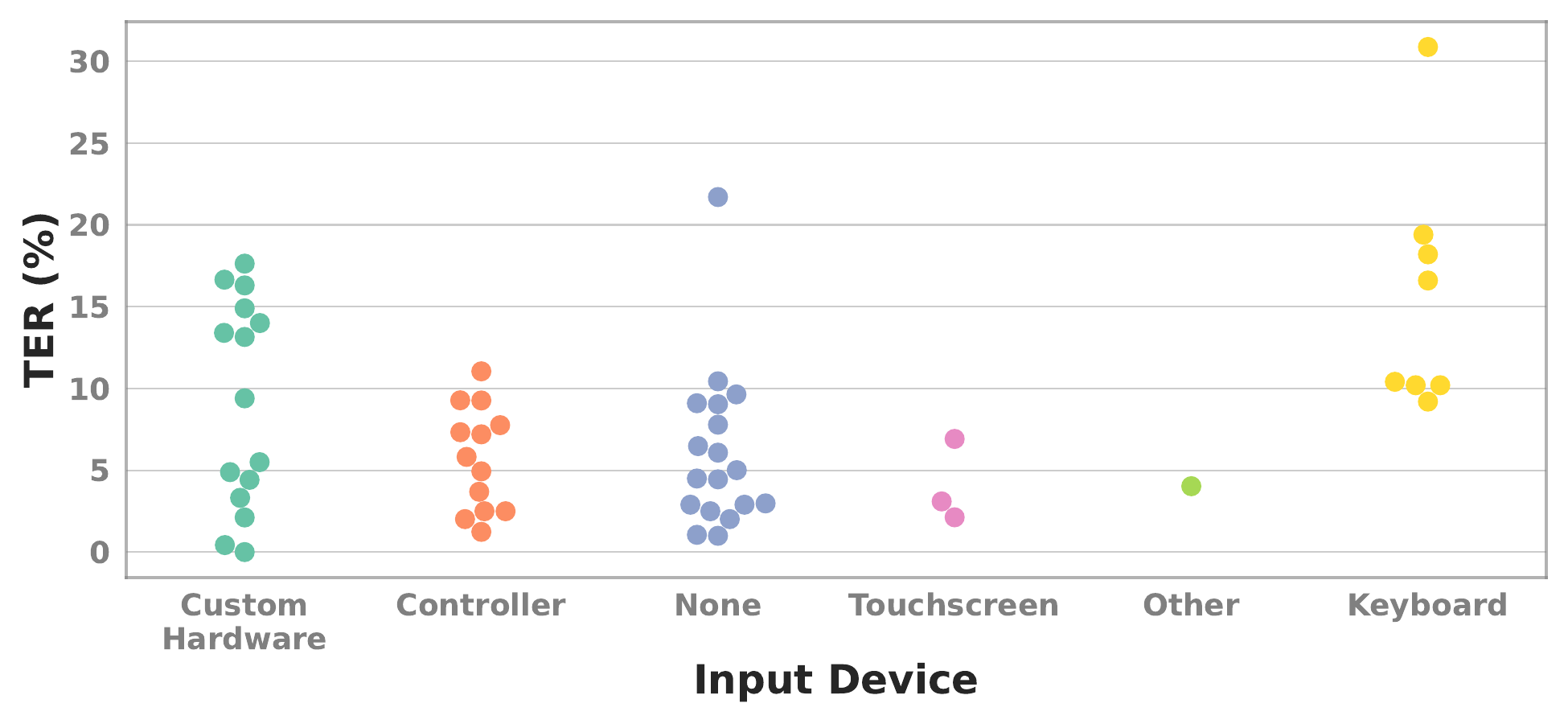}
            \label{fig:atribute_vs_metric:inputdevice_TER}
        \end{subfigure}
        \begin{subfigure}[b]{\textwidth}
        \caption{}
            \includegraphics[width=\linewidth]{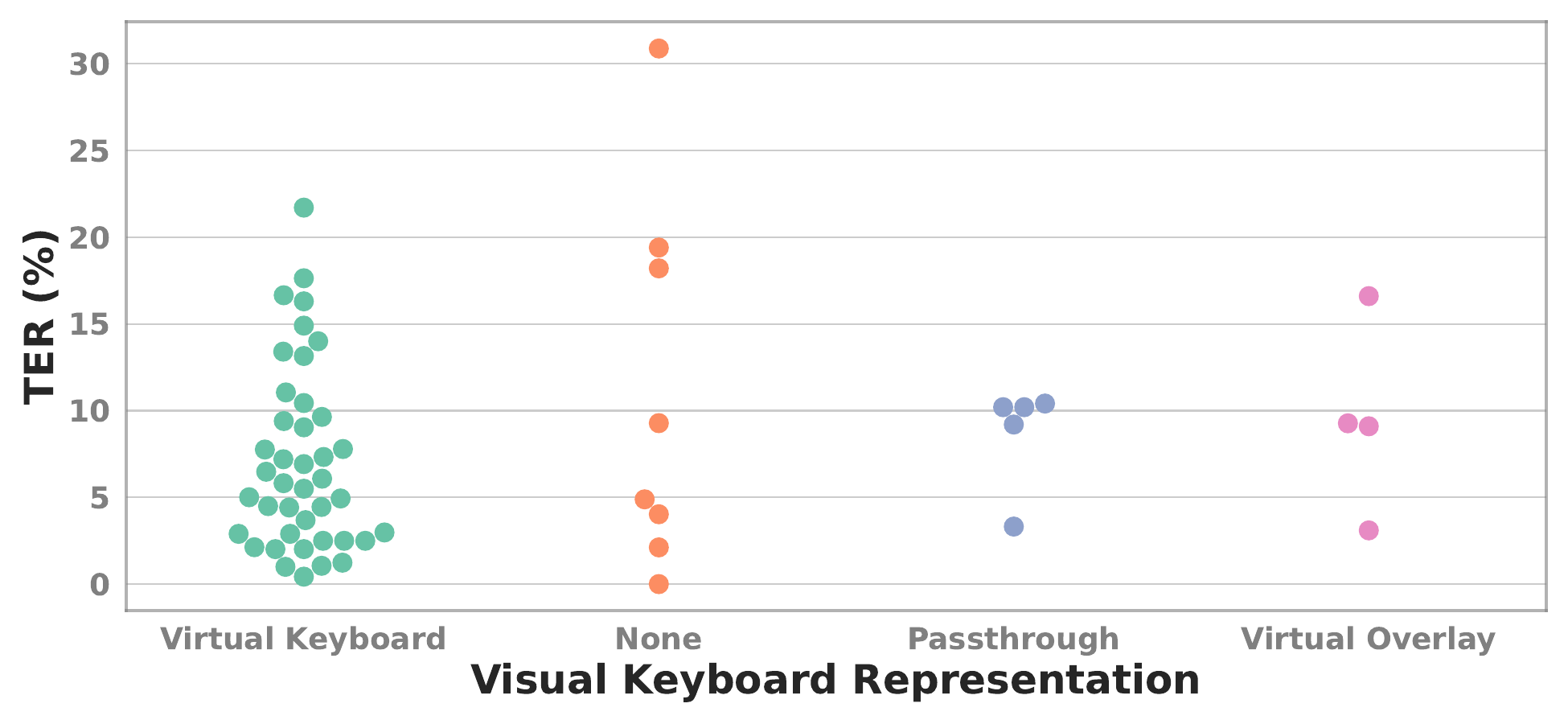}
            \label{fig:atribute_vs_metric:bodypart_TER}
        \end{subfigure}
    \end{minipage}
    \caption{WPM and TER by attribute (TETs having two attributes are plotted twice)}
    \label{fig:atribute_vs_metric}
    \Description{Figure 4 contains 3 scatter plots about words per minute versus concurrency (4a), input device (4b), and visual body representation (4c) and 3 scatter plots about Total error rate versus concurrency (4d), input device (4e), and body part attributes (4f).
    
    Figures 4a and 4d show the word per minute and total error rate versus one, two, and multiple concurrencies. The concentration of the data points increases as the concurrency increases for both metrics.

Figures 4b and 4e show the words per minute and total error rate versus the input devices (Custom Hardware, None, Other, Controller, Keyboard, and Touchscreen). The figure portrays that keyboards consistently have higher typing speeds compared to other input devices that are more spread out at lower speeds, they also have the highest error rate.

Figure 4c shows the words per minute versus the visual body representations (Invisible, Bones, Rays or Drums, Cursor or Pointer, Full Hand, Fingertips, and Controllers). 4c shows that cursor or pointer and controllers are consistently slower than other representations. Full-hand representations vary between 10 and 40 words per minute. The faster techniques were using fingertips around 60 words per minute.

Figure 4f shows the total error rate versus the body part (Finger, Hand, Head, Gaze). 4f shows that gaze and head have the least errors and fingers have the highest.
    }
\end{figure*}

\subsubsection{Typing Speed}
In our database, 137 techniques (133 + 4 converted from CPM) have typing speed data in terms of Words Per Minute (WPM) which we used to build our model. 
The features with greater than 5\% Gini importance were: Concurrency: multiple (0.4183), Input Device: physical Keyboard (0.0806), Concurrency: one (0.0521), and Visual Feedback: hovering (0.0504).

This list tells us which features impact typing speed the most but not how they impact it. To understand this, we look at patterns in our data. Concurrency (Figure \ref{fig:atribute_vs_metric:concurrency_WPM}) is the most important attribute for increasing typing speed which contributes by allowing multiple pointers that enable faster selection. Physical keyboards (Figure \ref{fig:atribute_vs_metric:inputdevice_WPM}) are the most commonly used text entry devices due to their efficiency. Having visual feedback for hovering allows easier key identification and helps further contribute to the effectiveness of concurrency by helping plan the next selection.

\subsubsection{Accuracy}
As discussed previously, 
accuracy is unfortunately not reported using the same metric for each technique. The error metrics we were able to obtain the most data for were character level TER (50) and character level MSD ER (44). Though we also had enough data for CER and UER, we decided to only focus on TER as it captures both values being their sum. Unfortunately, the subset of techniques that reported MSD ER values was not capturing enough of the attributes and had only a single technique for a lot of the attribute values, and thus we did not build a model for this metric.

For TER, the features having greater than 5\% Gini importance were: Concurrency: one (0.2525), Visual body representation: invisible (0.0746), Visual keyboard representation: None (0.062), Input Device: physical keyboard (0.0588), and Concurrency: multiple (0.0568).

Separating features corresponding to high and low TER values, we find that lower concurrency leads to lower error rates (Figure \ref{fig:atribute_vs_metric:concurrency_TER}), which is caused by minimization of input overlap and reduction in cognitive load. Surprisingly, physical keyboards have high TER values in our dataset compared to other input devices (Figure \ref{fig:atribute_vs_metric:inputdevice_TER}). This may be due to the closer proximity of keys which increases the chance of pressing incorrect keys. Not being able to see our body or the keyboard while typing causes higher error rates due to not having any visual feedback for positioning our body for correct input. 



\begin{figure*}[t]
    \centering
    \begin{subfigure}[b]{0.48\textwidth}
        \centering
        \caption{}
        \includegraphics[width=\textwidth]{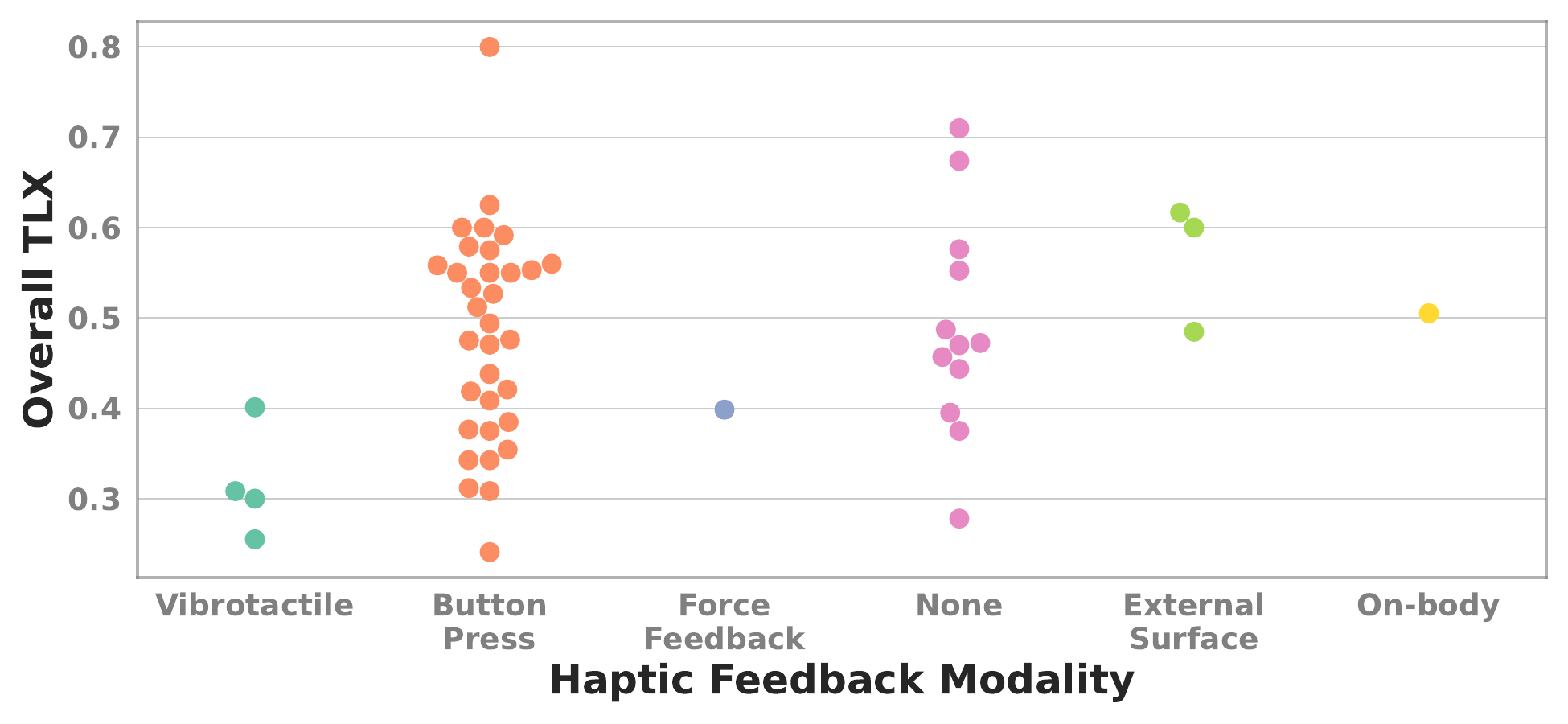}
        
         \label{fig:TLX_vs_attributes:TLX_vs_hapic}
    \end{subfigure}
    \hfill
    \begin{subfigure}[b]{0.48\textwidth}
        \centering
        \caption{}
        \includegraphics[width=\textwidth]{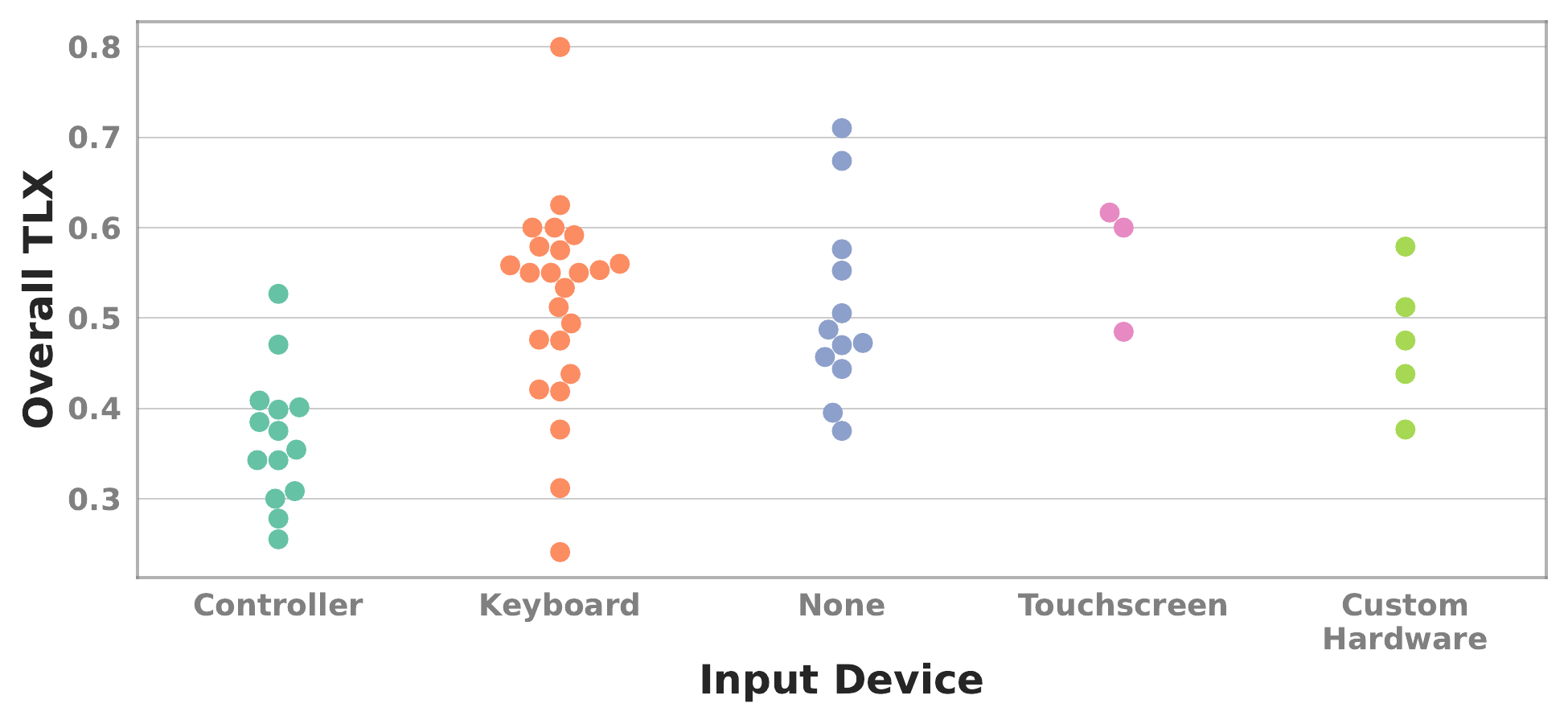}
        
         \label{fig:TLX_vs_attributes:TLX_vs_inputdevice}
    \end{subfigure}
    \vfill
    \begin{subfigure}[b]{0.48\textwidth}
        \centering
        \caption{}
        \includegraphics[width=\textwidth]{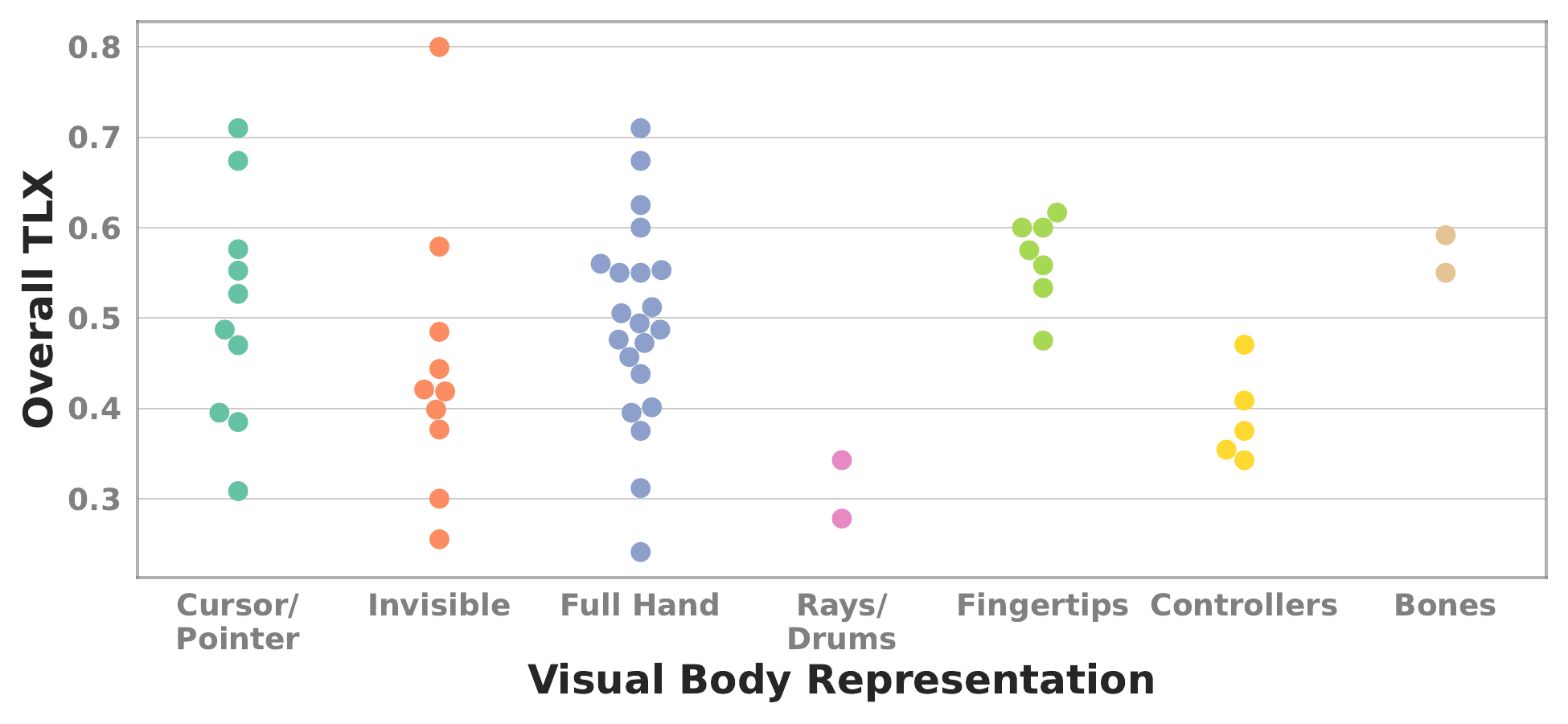}
        
         \label{fig:TLX_vs_attributes:TLX_vs_bodyrep}
    \end{subfigure}
    \hfill
    \begin{subfigure}[b]{0.48\textwidth}
        \centering
        \caption{}
        \includegraphics[width=\textwidth]{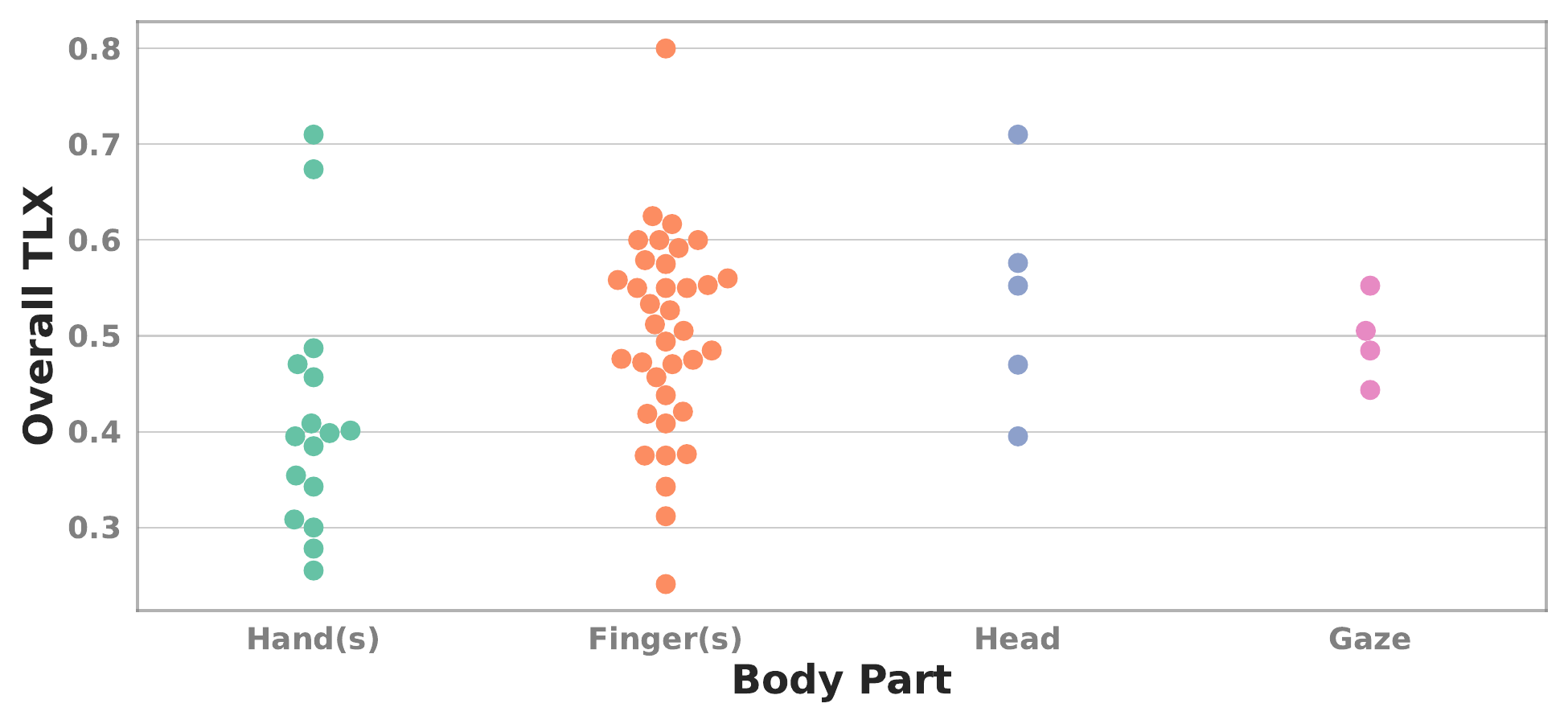}
        
         \label{fig:TLX_vs_attributes:TLX_vs_bodypart}
    \end{subfigure}
    \caption{Overall TLX by attribute (TETs having two attributes are plotted twice)}
    \label{fig:TLX_vs_attributes}
    \Description{Figure 5 contains 4 scatter plots of Overall TLX versus Haptic feedback modality, Input device, Visual body representation, and body part.}
\end{figure*}

\subsubsection{NASA TLX Scores}
We were able to collect or infer overall NASA TLX scores for 46 entries in our database. 
Not every study reports values for all six subscales and hence we do not create separate models for every subscale. Since studies used questionnaires with different point scales, we normalized the scores to be between 0 and 1. The features related to the NASA TLX score with greater than 5\% importance in the model were: Input Device: Controller (0.2765), Body part for input: hand(s) (0.1104), Body part for input: head (0.0599), Visual body representation: fingertips (0.0558), and Haptic feedback modality: vibrotactile (0.0504).

Separating the ones corresponding to high and low values, we see that controllers (Figure \ref{fig:TLX_vs_attributes:TLX_vs_inputdevice}) correspond to lower scores, reflecting their reduced workload across mental, physical, and temporal domains from only entering one or two characters at a time with small movements. For similar reasons, using hands as the body part for input (Figure \ref{fig:TLX_vs_attributes:TLX_vs_bodypart}) corresponds to lower scores. Using the head for input leads to higher TLX scores due to putting extra strain on the neck, head, and even upper body muscles. Having fingertips as the visual body representation causes higher TLX scores, perhaps since it is challenging to visually track and make sense of multiple small spheres (as opposed to hands or a single pointer) as one types. 
leading to higher mental and physical demands. Having vibrotactile feedback leads to lower scores as tactile confirmations of each keystroke have been shown to reduce the number of extra hand movements as well as failure to activate a key \cite{maunsbach2022whole,oulasvirta2018neuromechanics}, thus leading to lower physical demand and frustration.

\begin{figure*}[!ht]
    \centering
    \begin{minipage}[t]{0.4\textwidth}
        \centering
        \begin{subfigure}[b]{\textwidth}
            \caption{Number of TETs Proposed Each Year}
            \includegraphics[width=\linewidth]{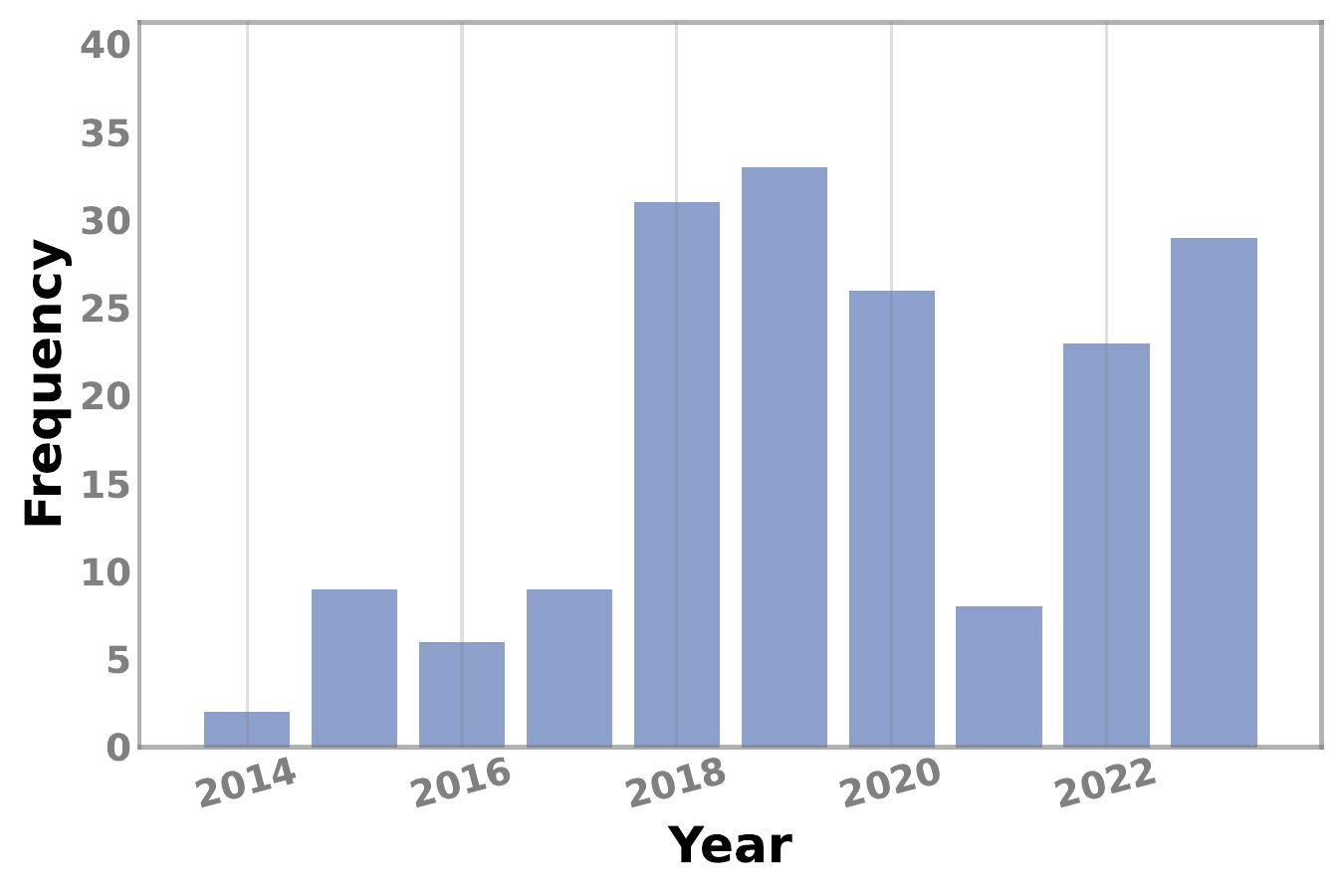}
            \label{fig:timetrends:Techniques}
        \end{subfigure}
        \begin{subfigure}[b]{\textwidth}
            \caption{Input Device Time Trend}
            \includegraphics[width=\linewidth]{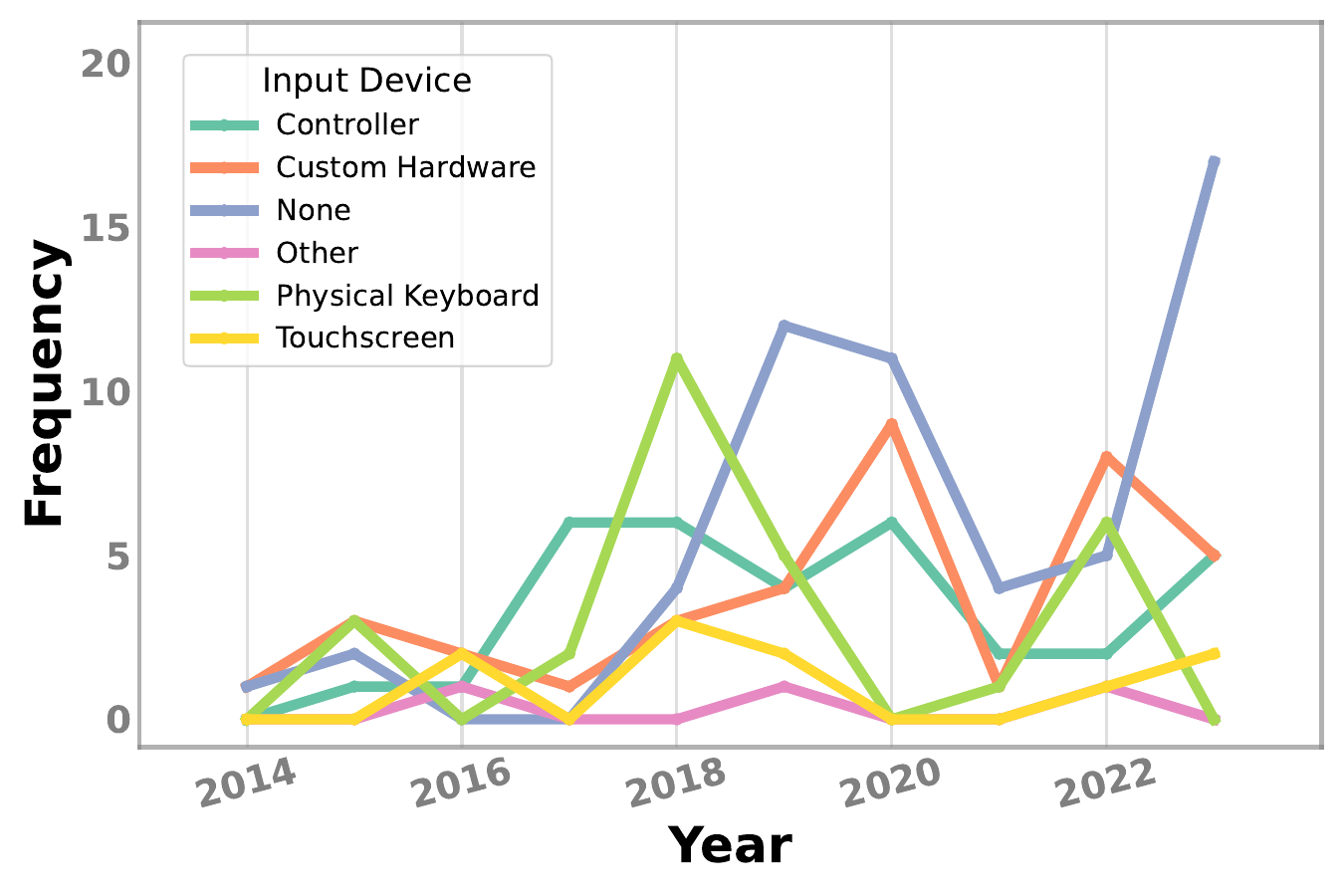}
            \label{fig:timetrends:Input_Device}
        \end{subfigure}
        \begin{subfigure}[b]{\textwidth}
            \caption{Concurrency Time Trend}
            \includegraphics[width=\linewidth]{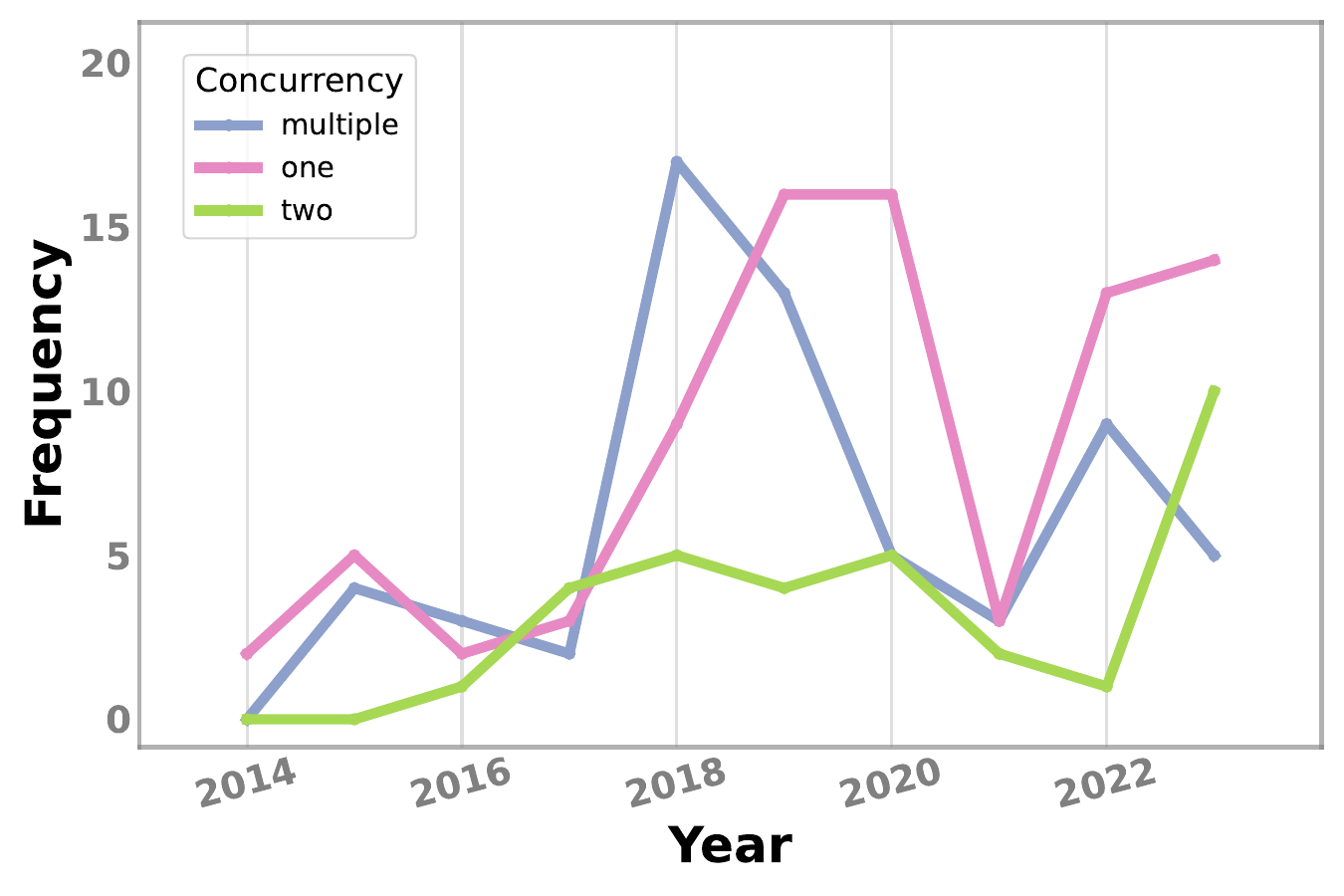}
            \label{fig:timetrends:Concurrency}
        \end{subfigure}
    \end{minipage}
    \hspace{4mm}
    \begin{minipage}[t]{0.4\textwidth} 
        \centering
        \begin{subfigure}[b]{\textwidth}
            \caption{Average WPM Each Year}
            \includegraphics[width=\linewidth]{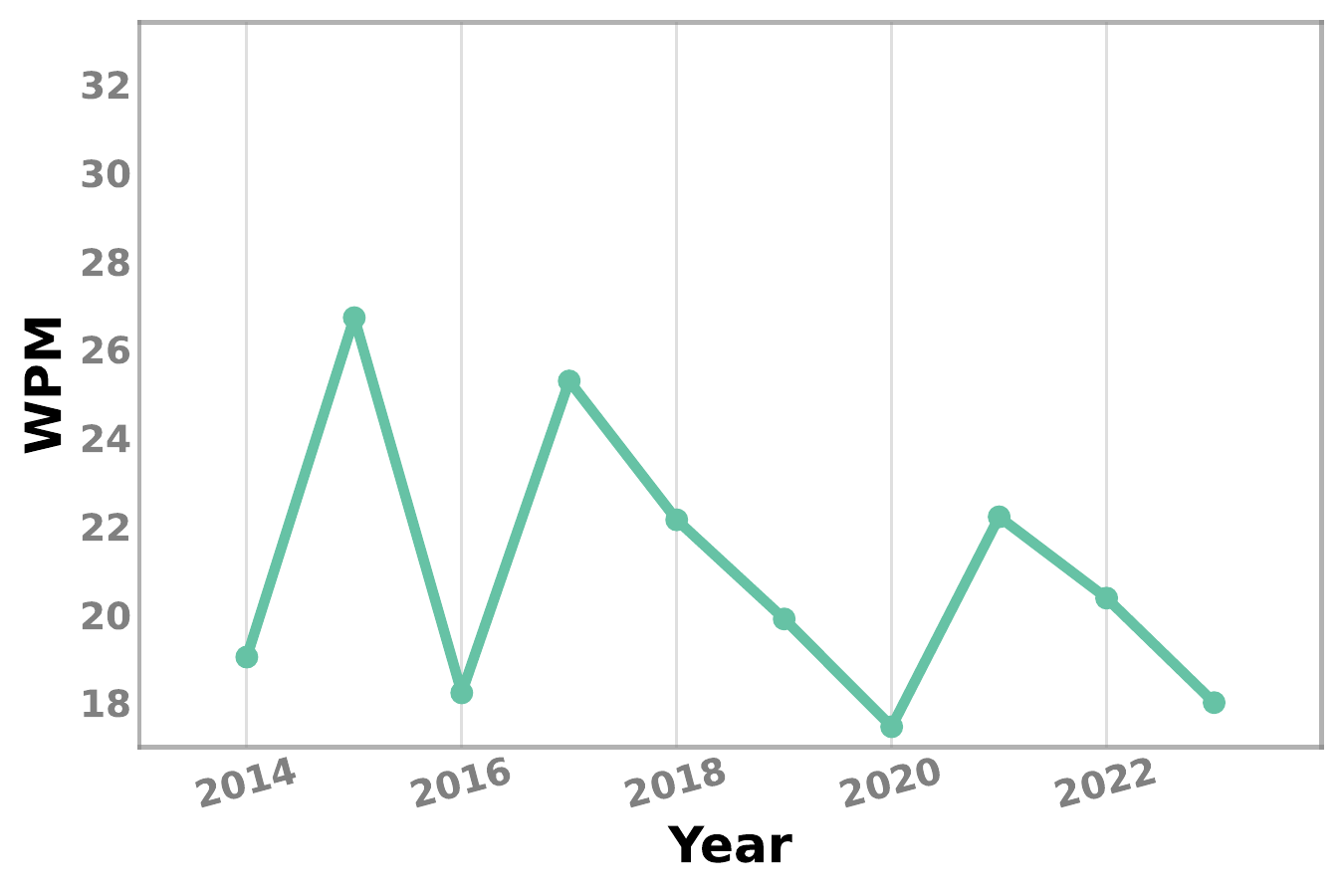}
            \label{fig:timetrends:WPM}
        \end{subfigure}
        \begin{subfigure}[b]{\textwidth}
            \caption{Average TER Each Year}
            \includegraphics[width=\linewidth]{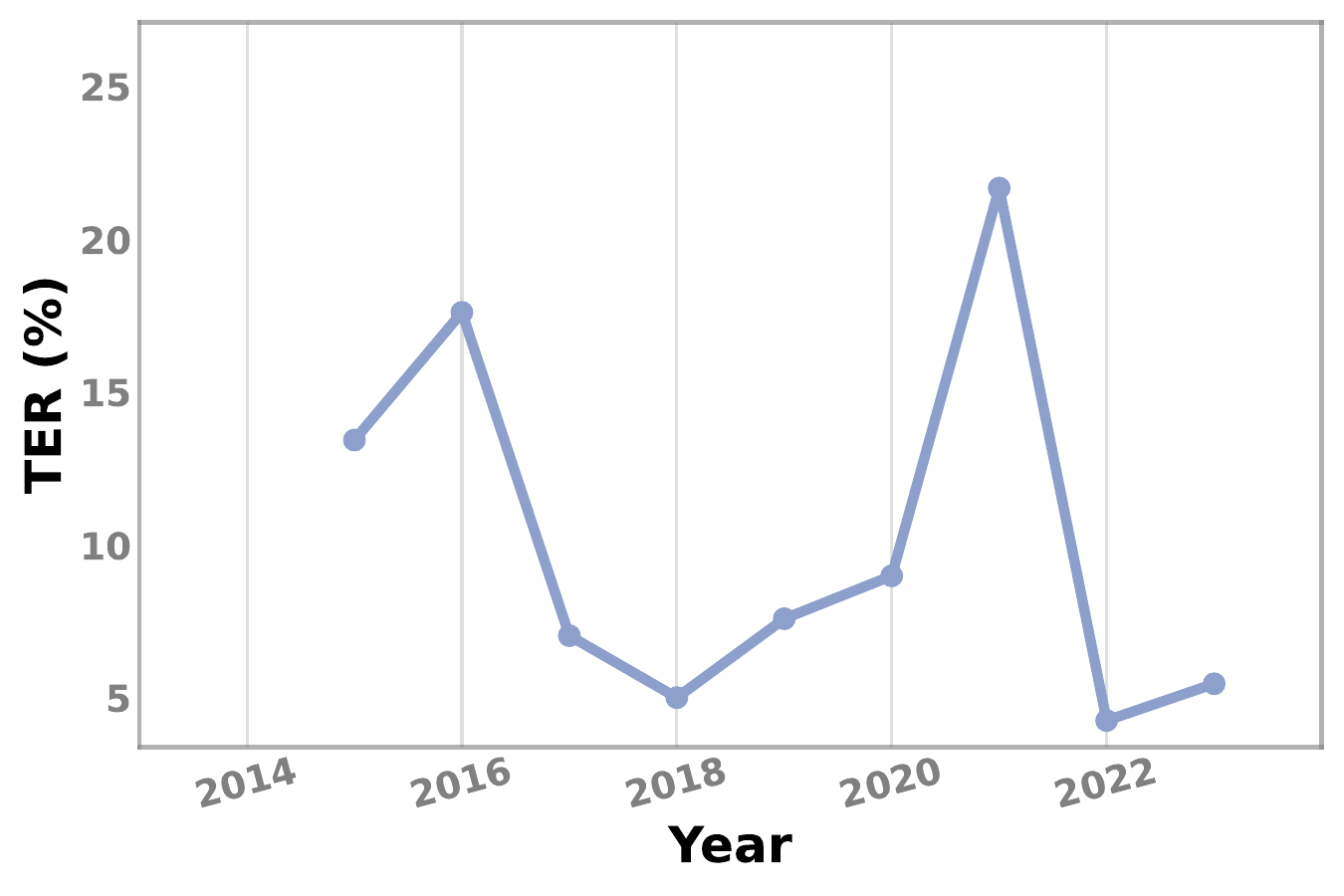}
            \label{fig:timetrends:TER}
        \end{subfigure}
        \begin{subfigure}[b]{\textwidth}
            \caption{Average MSD ER Each Year}
            \includegraphics[width=\linewidth]{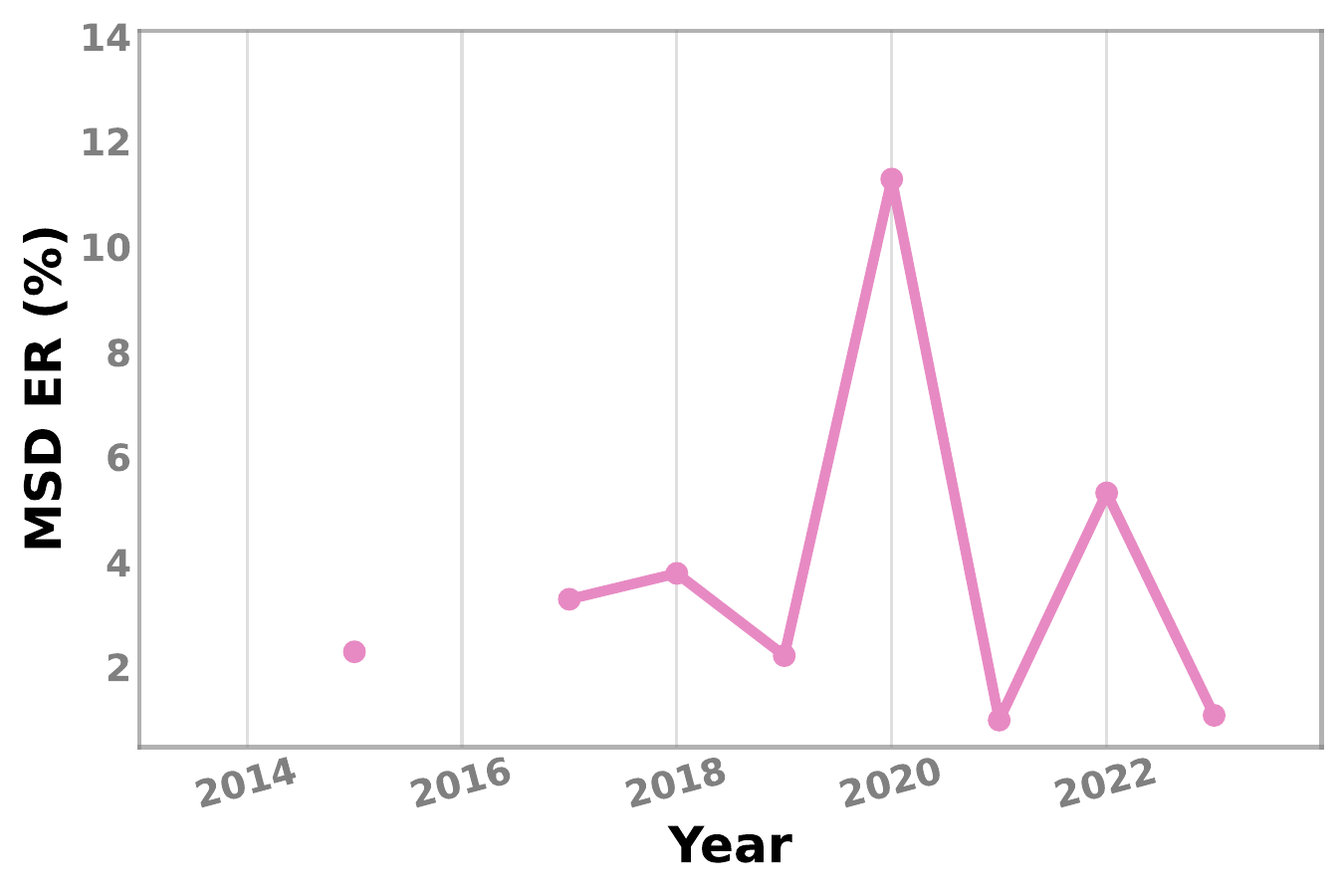}
            \label{fig:timetrends:MSDER}
        \end{subfigure}
    \end{minipage}
    \caption{Trends Over Time for Text Entry Techniques for XR}
    \label{fig:timetrends}
    \Description{Figure 6a shows an increase in the number of proposed metrics in 2018 to 2019 then a dip in 2021 then a recent increase.

Figure 6b input device shows a spike in keyboard TETs in 2018, and another spike in 2019 till 2021 then a recent interest in 2022 and 2023 for none as the input device.

Figure 6c increase in multiple and one as the concurrency in 2018 to 2021. Two stays constant and increases in 2023 with one.

Figure 6d shows 3 spikes every 2-3 that have lower peaks over time. Words per minute decrease over time.

Figure 6e shows a dip in the Total error rate between 2017 and 2020 then a spike in 2021 with a sudden decrease in 2022.

Figure 6f shows a spike in the minimum string distance error rate in 2020 that decreases in 2021.}
\end{figure*}

\subsection{Trends Over Time}

Figure \ref{fig:timetrends} shows time trends for TETs, reflecting advancements in commercial XR technology. Notably, there was a spike in techniques proposed in 2018 after the release of the HTC Vive and Oculus Rift and Touch controllers in 2016 and another spike in 2023 following the Meta Quest Pro release. For most trends, there is a drop in 2021, likely due to COVID-19. Below we further analyze the time trends for the interaction attributes and performance metrics to assess the field's progress and its relation to the importance of TET attributes.
 

\subsubsection{Input Device and Concurrency (Figures~\ref{fig:timetrends:Input_Device},~\ref{fig:timetrends:Concurrency})} We identified several trends in the \textit{input device} attribute, which are primarily influenced by commercial XR controllers and HMD technology. The use of ``none'' or no input device for text entry has risen since 2018, peaking in 2023, likely due to affordable hand-tracking tech like the Leap Motion and its VR SDK in 2016, and Meta Quest's improved hand-tracking in 2019. ``Physical keyboards'' saw a spike in 2018, possibly due to the HTC Vive Pro's pass-through camera but lowered in recent years. Controller use peaked in 2017 and has remained steady. Custom hardware for input has been among the top 2-3 trends over the years, suggesting the continuing need for XR hardware innovation for typing. In terms of \emph{concurrency}, ``multiple'' inputs increased in 2018 with the use of physical keyboards, but most techniques still focus on ``one'' input. As we showed above, concurrency is key to typing performance. Thus, increasing concurrency is an important area for further research.


\subsubsection{Typing Speed and Accuracy (Figures~\ref{fig:timetrends:WPM},~\ref{fig:timetrends:TER},~\ref{fig:timetrends:MSDER})} Surprisingly, the average input speed has a downward trend over the years, likely due to reduced concurrency as new techniques adapt to commercial XR hardware rather than developing custom text input devices. 
Regarding \emph{accuracy}, error rates have slightly decreased over time, possibly due to improved tracking and error correction algorithms such as those for reducing co-activations \cite{10.1145/3411764.3445671}. The spike in MSD ER in 2020 is attributed to the pen-based Arc-type technique\cite{10.1145/3385959.3418454}, which required very little movement but had the highest error rate in the database due to jitters. The increase in TER in 2021 reflects limited TER data that year due to the pandemic.


\subsubsection{Other Trends} Other trends are related to body part, visual body representation, keyboard representation, and feedback. Fingers, followed by hands, have been the most common input methods, likely due to advancements in VR controllers and hand tracking since 2016. Visualizing the full hand increased in 2023, possibly due to pass-through cameras, while the use of cursors or pointers was more common from 2017--2020. Our analysis suggests that visualizing full hands can increase the error rate (TER), and thus should be considered with caution in future TETs. Virtual keyboards have been consistently frequent, with virtual overlays peaking in 2018--2019 but fading afterward. 
For haptic and auditory feedback, button presses were common in 2018-2019, but there's been a rise in ``on-body,'' ``external surface,'' and ``none'' feedback in 2022-2023. Not surprisingly, feedback is often for ``key activation'' or ``none'' at all. 



\section{Discussion}

This work focused on collecting and analyzing 176 existing text entry techniques (TETs) in XR and creating a tool to navigate this space. Our interactive tool aims to support XR developers in selecting or designing appropriate techniques for their applications. While we analyze attributes and metrics of TETs, XR use cases are diverse, making it impractical to define a single TET solution or categorization that is suitable for every application. Instead, our online TEXT\footnote{{https://xrtexttrove.github.io/}} interface aims to support designers to efficiently navigate the design space of techniques and filter them by their interaction characteristics, use cases (e.g., on-the-go, AR), or target performance to identify a subset of candidate techniques to test for their applications. 

Below, we present takeaway findings for XR designers and researchers, reflect on the implications of rapid progress in augmented realities for XR text entry, and discuss progress in the related area of text editing in XR.



\subsection{Implications for Designing New XR Text Entry Techniques}

For future works in text entry for XR, we provide the following recommendations based on our analysis of existing techniques. 

\subsubsection{Focus on building for text entry rather than adapting for it}
Many techniques have been proposed since 2018, suggesting a demand for new solutions in this area. The evolution of these techniques has mirrored trends in commercial XR technologies, such as the rise of no-input devices with advanced hand tracking. However, metrics such as typing speed have not improved over the years. Developing custom text input devices and software optimizations for XR, such as the TapXR\footnote{{https://www.tapwithus.com/}} or the decoder by \citet{8943750}, should be explored further for improvements to typing performance. Identifying and attempting to fix bottlenecks in current technologies such as bad quality pass-through cameras could be another way to improve performance.

\subsubsection{Focus on concurrency and input device}
Our analysis suggests that concurrency and input device are the most important interaction attributes for all three measures, speed, accuracy, and task load. Concurrency is key for improving typing speed, yet most techniques have focused on interactions that afford one or two inputs. Though unable to retain the same performance inside XR, keyboards still enable high-speed typing albeit with more errors.  The high speed is perhaps due to user familiarity and kinesthetic learning of button positions and physical feedback with these keyboards. The higher error rate can be due to the close placement of keys in these keyboards and the lack of effective mechanisms (e.g., passthrough) to accurately visualize the keyboards until recently. Controllers are consistently slow but offer low error rates and cognitive load, perhaps due to their single concurrency input, improvements in their ergonomic design over years, and physical button feedback.  
Other technologies, such as gaze tracking (i.e., input modality: gaze), did not show a significant trend in our data, perhaps because these technologies are still improving and thus are less explored for text input by XR designers. 
Unlike laptops and personal computers, no commercial XR headset currently comes packaged with a text input device despite its importance. This will change in the future if we develop text entry devices specifically for XR, as mentioned above. Beyond concurrency and input devices, visualizing full hands or fingertips can improve typing speed but also increase error rates and workload.

\subsubsection{Use standardized measures for evaluation, namely WPM, TER, and NASA TLX}
Our findings further highlight the lack of standards for reporting and benchmarking TETs. 
The inconsistency in reporting error rates hinders comparison for designers and researchers. We faced this issue when predicting the impact of interaction attributes on error rates and had to conduct separate analyses for subsets of TETs. When \citet{SoukoreffMackenzie2003ErrorMetrics} introduced TER as a measure of errors, they noted that it encapsulates error better than MSD ER \cite{SoukoreffMackenzie2003MSDER}. A later review on performance metrics for text entry also identified TER as ``the most powerful error metric'' because it combines both persistent and corrected errors \cite{ArifStuerzlinger2009}. Following the previous findings, we recommend new TETs to include TER as one of their reported error metrics and clearly distinguish whether they use character-level or word-level TER in reporting the error value. Similar inconsistencies are present for task load, where there is a need to use standardized questionnaires instead of creating custom ones. The phrase sets used for evaluation must also be chosen from one of the standardized corpa such as the MacKenzie and Soukoref phrase set \cite{MacKenzieSoukoreffPhraseSet2003} or the Enron mobile dataset \cite{EnronPhraseSet2011}. This is because these phrase sets cover real-world scenarios and are widely adopted which ensures consistency across studies allowing easier comparisons and reproducibility. A potential solution is the creation of a tool like TextTest++ \cite{Zhang:2019:BIS:3332165.3347922} for XR that provides a standardized environment for text entry and automatically calculates the standard metrics.

\subsubsection{Evaluate fatigue}
Ergonomics and comfort are important issues in XR interaction techniques \cite{10.1145/3411764.3445349, 10.1145/3411764.3445361}, especially those involving prolonged tasks such as text input. Unfortunately, very few techniques are evaluated on this factor, and those that do often use custom questionnaires. Future techniques should use standardized questionnaires such as the Borg CR10 scale \cite{borg1985increase}.

\subsubsection{Account for Learning Effects}
Study design can affect the text entry performance if the participants get better with repeated trials. Thus, to be able to compare techniques, studies must ensure the users pass the initial learning curve to get the most accurate assessment of a technique's performance. In our review, we categorized whether the technique was evaluated in a single or multi-session study. Only 20\% (N=29) of techniques that conducted user studies (N=140) leveraged a longitudinal design.  While multi-session studies are not essential for every technique, novel techniques must carefully measure the users' learning rate and include multi-session studies as needed to accommodate for learning effects. 


\subsubsection{Design for context and use it for evaluation}
Techniques should be specifically designed for the context of their use, such as in an office, texting on the go, or in social settings. While some researchers have optimized text entry for certain use cases, such as privacy \cite{8794572}, mobility \cite{Rajanna2018onthegostudy1}, and accessibility \cite{10.1145/3380983}, most techniques do not discuss a context. Achieving higher performance is meaningless if it cannot be achieved when used in the actual setting of an application. Techniques should not just be designed for contexts but also evaluated in them. If a technique is optimizing for certain parameters such as fatigue or mobility, it should be evaluated in relevant contexts such as for writing multiple pages of text or walking while encountering obstacles. 




\subsection{Adoption of Augmented Reality}
Augmented reality technology has improved by a large margin in the last couple of years with good quality video see-through displays available such as the Apple Vision Pro and Meta Quest 3. Having better pass-through cameras with less distortion and lag can improve the performance of pass-through-based techniques which currently suffer high error rates. Better tracking capabilities can improve keyboard backends based on the understanding of the user and their surroundings.

With augmented reality becoming ubiquitous, the design and evaluation of TETs would then also need to consider the context of use, something that is rarely investigated today. Being able to use XR anywhere may involve frequent switching of TETs or modification of certain attributes based on the context of use. 
These switches need to be as seamless as possible, hence learnability and the cost of switching between TETs would be important metrics to investigate. 
With AR being employed in public spaces, the presence of bystanders while performing text entry has important implications on the selection and appeal of a TET. In particular, the social acceptability of a technique needs to be considered and evaluated.
Measures involving text intelligibility and security \cite{8794572} would also need to be developed for cases where information being typed needs to be protected.



\subsection{Text Selection and Editing Techniques in XR}
The challenge of text entry is that it normally doesn't happen in isolation. To be truly meaningful, text entry must enable actual work and productivity. That means that text entry comes together with text editing, with the capacity to delete, select, copy, paste, and to move the cursor within the text into a particular spot to introduce changes and comments or format the text. 
These additional high-precision input tasks must be considered together with text entry \cite{gonzalez2024guidelines}. 

In that regard, some prior work has already approached the problem from a holistic perspective, for example, exploring how to do text revision with backspace and caret in virtual Reality \cite{10.1145/3411764.3445474}
or evaluating caret navigation methods for text editing in augmented Reality \cite{9974482}. This is a particular challenge for any voice-enabled text entry too, as oftentimes the input needs to be updated. For selecting text \citet{9995155} have looked at hands-free selection methods in virtual reality. Compared to text entry, text editing in XR is still a nascent field and can benefit from the findings for text entry as well as general selection techniques in XR \cite{10.1145/3411764.3445193}.


\section{Limitations and Future Work}

Our work is constrained by inconsistencies in reporting performance metrics for TETs and the various study designs and setups that can impact TET performance. 
Thus, we identify trends in the data and factors typically associated with high-performing techniques and do not directly run statistical analysis between the studies. For the same reason, we do not predict values for the performance of a technique or determine the best technique within our dataset. To address this limitation, one could run a large-scale crowdsourcing study of TETs to provide a direct comparison and address the missing data in our trove. Such a study can include new techniques and variations of existing ones to create a test set with standardized performance measures and further assess the generalizability of trends reported here. Another approach could involve modeling the underlying relationships between different metrics through methods like symbolic regressions \cite{reinbold2021robust} or simulation studies of human motor control and typing \cite{ikkala2022breathing,shi2024crtypist}
to estimate a missing metric from reported values or user interactions. While we made an early attempt at this, our search did not result in a validated formula, making such an investigation an open area for future work. 

Relatedly, for identifying attribute importance, we used Gini importance which is known to have limitations such as sensitivity to correlations between features and bias toward high-cardinality features \cite{strobl2007bias}. To mitigate this, we 
only report those attributes with importance greater than 5\% in this paper. With a larger dataset, future work can compare our results against outcomes of other statistical techniques such as permutation importance \cite{terence2018beware} to provide further insights into the importance of interaction features and improve generalizability for unseen data. 

The dataset created in this work is meant to be comprehensive but not exhaustive. There may be techniques we missed that were not published in mainstream venues or created by hobbyists but not shared on social media. We hope that the suggestion form feature of the TEXT tool can help further increase the size of our dataset and keep it updated over time. Similarly, the set of attributes to describe the techniques capture a majority of the variations present in existing techniques but do not capture every possible difference such as details about hand pose during typing \cite{9737726} or location of haptic feedback \cite{gupta2020investigating}. These variations may be viable attributes to tweak when designing future techniques after optimizing for the existing important attributes identified by our work.

An issue common to datasets like the one presented here is maintaining them as new techniques appear \cite{seifi2019haptipedia,10.1145/3411764.3445319}. The challenge is that the dataset cannot be fully open to external updates, as this could compromise the quality of the labeling or other aspects of the dataset, and as such they often fall outdated. This problem might be solved in the future by leveraging large language models (LLMs). There are a number of synthetic labeling initiatives \cite{syntheticusersSyntheticUsers} and cases of AI producing high-quality labeling when provided with good examples and advanced prompts \cite{abreu2024parse}. Once verified, this approach could provide a sustainable way to update the dataset as new techniques emerge. 

\section{Conclusion}
In this work, we created a dataset of 176 text entry techniques (TETs) for XR from across academia, industry, and hobbyists. We described each technique in 13 interaction attributes, 14 performance metrics, and 5 general codes. We then created an online tool, TEXT: Text Entry for XR Trove to be able to visualize our dataset and navigate the solution space for text entry in XR. By analyzing our dataset, we highlight trends in the design of TETs, their evaluations, and the relative importance of interaction attributes when trying to optimize the performance of a technique. This work is a step towards future XR productivity tools that enhance user performance and experience beyond the physical keyboard. 

\begin{acks}
This work was supported by research grants from Google, the Max Planck Institute for Intelligent Systems, VILLUM FONDEN (VIL50296), the Engineering and Physical Sciences Research Council (EP/V034987/1) and the National Science Foundation (2339707). We thank Tor-Salve Dalsgaard for providing the script to filter results from the Springer website.
\end{acks}

\bibliographystyle{ACM-Reference-Format}



\nocite{*}
\end{document}